\documentclass[amssymb,11pt]{article}
\usepackage{amsmath}
\usepackage{amsmath}
\usepackage{amssymb}
\usepackage{graphicx}

\def\>{\ensuremath{\rangle}}
\def\<{\ensuremath{\langle}}

\newtheorem{thm}{Theorem}[section]

\newtheorem{defn}{Definition}[section]
\newtheorem{prop}{Proposition}[section]
\newtheorem{exam}{Example}[section]

\usepackage[ruled]{algorithm2e}

\SetAlFnt{\small}
\SetAlCapFnt{\small}
\SetAlCapNameFnt{\small}
\SetAlCapHSkip{0pt}
\IncMargin{-\parindent}

\usepackage{times}
\usepackage{amsmath}
\usepackage{stmaryrd}
\usepackage{amssymb}

\def\>{\ensuremath{\rangle}}
\def\<{\ensuremath{\langle}}

\begin{document}
\title{Quantum Information-Flow Security: Noninterference and Access Control}
\author{Mingsheng Ying, Yuang Feng and Nengkun Yu\\
\\
\small \em QCIS, FEIT, University of Technology, Sydney, 
Australia\\
\small \em and\\
\small \em TNList, Dept. of CS, Tsinghua University, China\\
\small \em Email: Mingsheng.Ying@uts.edu.au, yingmsh@tsinghua.edu.cn}
\date{}
\maketitle
 
\begin{abstract}
Quantum cryptography has been extensively studied in the last twenty years, but information-flow security of quantum computing and communication systems has been almost untouched in the previous research. Duo to the essential difference between classical and quantum systems, formal methods developed for classical systems, including probabilistic systems, cannot be directly applied to quantum systems. This paper defines an automata model in which we can rigorously reason about information-flow security of quantum systems. The model is a quantum generalisation of Goguen and Meseguer's noninterference. The unwinding proof technique for quantum noninterference is developed, and a certain compositionality of security for quantum systems is established. The proposed formalism is then used to prove security of access control in quantum systems.  
\end{abstract}

\section{Introduction}

It is well-known that quantum cryptography has a great advantage over its classical counterpart that the security and ability to detect the presence of eavesdropping are provable, based on the
principles of quantum mechanics. But it has been rarely noticed that quantum computing and communication systems also face a new security challenge that would not arise in classical systems:   
entanglement is indispensable in quantum computation and communication, but information leakage can be caused by an entanglement (or more precisely, a computational mechanism that can generate an entanglement, e.g. the CNOT gate; see Examples~\ref{ex1} and~\ref{ex2}), and thus the Trojan Horse may exploit an entanglement between itself and a user with sensitive information as a covert channel.  

Information-flow security policies are usually enforced to prevent improper information leakage in classical computing and communication system~\cite{SM03}.  
A general framework for specifying and analysing information-flow security is the noninterference formalism first introduced by Goguen and Meseguer~\cite{GM82}. 
The basic idea of noninterference~\cite{GM82} is: \begin{itemize}\item 
 \textquotedblleft One group of agents, using a certain set of commands, is non-interfering with another group of agents if what the first group does with those commands has no effect on what the second group can see."  
\end{itemize} Then information leakage from a group of agents to another group of agents is understood as interference of the first group with the second group, and security is defined as noninterference of the agents with sensitive information with those malicious agents. 
In the original formulation~\cite{GM82} of noninterference, its system model is a deterministic automaton. This model has been generalised to a nondeterministic automaton by Sutherland~\cite{Su86} and McCullough~\cite{Mc88} and further to a probabilistic automaton by Grag~\cite{Gr90}. 

This paper aims at extending further the noninterference formalism so that it can be used to reason about information-flow security of quantum systems. A quantum system is in a sense a probabilistic system, but the theory of probabilistic noninterference~\cite{Gr90} cannot be directly applied to it due to the following two reasons: \begin{enumerate}\item In a quantum system a probability distribution of outputs only appears after a certain measurement. Any observation about a classical or probabilistic system by an agent does not disturb the state of the observed system and thus has no interference with other agents. However, a basic postulate of quantum mechanics stipulates that the only way for acquiring information about a quantum system is quantum measurement, which will alter the state of the observed system. Thus, interference between different agents will be introduced during observation on quantum systems.  
\item The computational steps of a quantum system are governed by unitary operators or more generally super-operators, which are essentially different from stochastic matrices that are commonly used to model the dynamics of probabilistic systems. In other words, the mathematical description of commands executed by an agent in a classical or probabilistic systems is different from that in a quantum system. 
\end{enumerate}  To appropriately incorporate quantum features into the noninterference formalism, we define a system model in terms of quantum automata~\cite{MC00}.   

Di Pierro, Hankin and Wiklicky~\cite{DHW02} observed that absolute noninterference can hardly ever be achieved in real systems, and thus they proposed a novel notion of approximate noninterference based on a quantitative measure of process behaviour equivalence. The non-appropriateness of absolute noninterference is even truer in the quantum case because quantum gates form a continuum and noise in their physical implementation is unavoidable. So, we define a quantitative version of noninterference (or approximate noninterference) for quantum systems, following Di Pierro, Hankin and Wiklicky~\cite{DHW02}. (A notion of approximate behaviour equivalence was also adopted by the authors in their work on both classical and quantum process algebras~\cite{YW00}, \cite{Yi01}, \cite{YFDJ09, FDY12}.)   

The main technical contribution of this paper are:
\begin{itemize}  
\item Unwinding proof technique: It is often hard to establish noninterference security because noninterference is defined as a property over sequences of commands of arbitrary length. A unwinding technique was proposed by Goguen and Meseguer~\cite{GM84}, which can prove noninterference by checking only certain single-step conditions. This technique was generalised by Rushby~\cite{Ru92} and van der Meyden~\cite{vdM07} to the case of intransitive noninterference. We further generalise this technique and provide a method for estimating the upper bound of insecurity degree of quantum system.
\item Compositionality of security: A research line on compositionality of security was initiated by McCullough~\cite{Mc88} and recently systemised by Mantel~\cite{Ma02}, 
showing that secure components with appropriate interface can be hooked up to form a secure system. As a quantum generalisation of their compositionality theorems, we prove that the insecurity degree of a composed quantum system does not exceed the sum of the insecurity degrees of their components provided no entanglement exists between those components. \end{itemize}

As an application of the proposed formalism, we consider access control of quantum data. The operating systems of all modern computers include certain form of access control to protect confidential data. Access control of quantum data will certainly be an important issue in the design of an operating system for future quantum computers. 
The simplest access control policy is usually defined in terms of access control matrix, which specifies the access rights of agents to individual storage locations. A quantum access control matrix is much more complicated than its classical counterpart due to a subtle difference between classical and quantum information: 
\begin{itemize}
\item \textquotedblleft$1+1<2$": Access to the quantum information stored in a composite $AB$ system is not granted by access to the information stored in subsystem $A$ and access to that in subsystem $B$ (see Example~\ref{112}).
\end{itemize} More precisely, a quantum access control matrix has to specify the access rights of agents not only to individual storage locations but also to different combinations of individual locations. Rushby~\cite{Ru92} showed by the unwinding technique that security of access control can be properly interpreted in the noninterference formalism with the Reference Monitor Assumptions. As a quantum generalisation of Rushby's result~\cite{Ru92}, we show that the insecurity degree of quantum access control is bounded by a linear function of the degree that the Reference Monitor Assumptions are satisfied. 

The paper is organised as follows. Since the majority of Computer Security Foundations community may have no background in quantum computation, we briefly review its basics including the mathematical formalism of the state space and dynamics of a quantum system and quantum measurement in Sec.~\ref{Dyn}; for more details we refer to \cite{NC00}. Another purpose of Sec. \ref{Dyn} is to fix notations used in the later sections. 
The automata model of quantum systems and a noninterference measure in such a model are introduced in Sec.~\ref{MOD}. In Sec.~\ref{SECT}, we define the core notion - security degree of quantum systems - in terms of the noninterference measure, and the unwinding technique for proving security is generalised to the quantum setting. A compositionality theorem for quantum security is established in Sec.~\ref{COMPS}. The security properties of access control of quantum data are examined in Sec. \ref{ACCC}. A brief conclusion is drawn in Sec.~\ref{concl}, including several problems for further research.  
For the readability, we postpone all the proofs of theorems to the Appendix.  

\section{Basics of Quantum Theory}\label{Dyn}

\subsection{Hilbert Spaces}

According to a basic postulate of quantum mechanics, the state space
of a quantum system is represented by a Hilbert space. In this paper, we only consider finite-dimensional Hilbert spaces, which are indeed complex vector spaces with inner product. We assume the reader is familiar with the notion of vector space in Linear Algebra. An inner product over a vector space
$\mathcal{H}$ is a mapping
$\langle\cdot|\cdot\rangle:\mathcal{H}\times \mathcal{H}\rightarrow
\mathbb{C}$ satisfying the following properties: \begin{enumerate}\item $\langle\varphi|\varphi\rangle\geq 0$ with equality if and only if
$|\varphi\rangle =0$; \item 
$\langle\varphi|\psi\rangle=\langle\psi|\varphi\rangle^{\ast}$; and
\item $\langle\varphi|\lambda_1\psi_1+\lambda_2\psi_2\rangle=
\lambda_1\langle\varphi|\psi_1\rangle+\lambda_2\langle\varphi|\psi_2\rangle$\end{enumerate} for any $|\varphi\rangle, |\psi\rangle, |\psi_1\rangle,
|\psi_2\rangle \in \mathcal{H}$ and for any $\lambda_1,\lambda_2\in\mathbb{C}$, where $\mathbb{C}$ is the field of complex numbers, and $^{\ast}$ stands for the conjugate of complex numbers. A vector $|\psi\rangle$ is called a unit vector if $\langle\psi|\psi\rangle=1$. 
A pure state of a quantum system is described by a unit vector in its
state space. Two vectors $|\varphi\rangle$ and $|\psi\rangle$ are said to be orthogonal, written $|\varphi\rangle\bot |\psi\rangle$ if $\langle\varphi |\psi\rangle=0$. A family $\{|\psi_i\rangle\}_{i=0}^{n-1}$ of unit vectors is called an orthonormal basis of $\mathcal{H}$ if \begin{enumerate}\item $|\psi_i\rangle\bot |\psi_j\rangle$ for any $i\neq j$; and \item $|\psi\rangle=\sum_{i=0}^{n-1}\langle\psi_i|\psi\rangle|\psi_i\rangle$ for all $|\psi\rangle\in\mathcal{H}$.\end{enumerate} In this case, $\mathcal{H}$ is said to be $n-$dimensional, each element $|\psi\rangle$ of $\mathcal{H}$ can be represented by a column vector $|\psi\rangle=(a_0,...,a_{n-1})^T,$ where $a_i=\langle\psi_i|\psi\rangle$ for $0\leq i<n$, and $^T$ stands for transpose.

\begin{exam}\label{ht} \textit{Quantum bit}, or \textit{qubit} for short, is the quantum counterpart of the bit in classical computation. The state space of qubits is the $2-$dimensional Hilbert space $$\mathcal{H}_2=\{\alpha |0\rangle+\beta|1\rangle:\alpha,\beta\in\mathbb{C}\}.$$ The inner product of $|\psi\rangle=\alpha|0\rangle+\beta|1\rangle$ and $|\varphi\rangle=\alpha^{\prime}|0\rangle+\beta^{\prime}|1\rangle$ is $$\langle\psi|\varphi\rangle=\alpha^{\ast}\alpha^{\prime}+\beta^{\ast}\beta^{\prime}.$$ 
The vectors $$|0\rangle=\left(\begin{array}{cc}1\\ 0\end{array}\right),\ |1\rangle=\left(\begin{array}{cc}0\\ 1\end{array}\right)$$ form an orthonormal basis of $\mathcal{H}_2$, called its computational basis. A qubit can be in the basis states $|0\rangle$ and $|1\rangle$ as well as their superpositions $$\alpha |0\rangle+\beta |1\rangle=\left(\begin{array}{cc}\alpha\\ \beta\end{array}\right)$$ where $|\alpha|^{2}+|\beta|^{2}=1$, such as \begin{equation*}\begin{split}|+\rangle&=\frac{1}{\sqrt{2}}(|0\rangle+|1\rangle)=\frac{1}{\sqrt{2}}\left(\begin{array}{cc}1\\ 1\end{array}\right),\\ |-\rangle&=\frac{1}{\sqrt{2}}(|0\rangle-|1\rangle)=\frac{1}{\sqrt{2}}\left(\begin{array}{cc}1\\ -1\end{array}\right).\end{split}\end{equation*}\hfill $\blacksquare$\end{exam}

The state space of a composite quantum system is defined to be the tensor product of the state spaces of its subsystems. Let $\mathcal{H}_i$ be a Hilbert space with $\{|\psi_{ij}\rangle\}$ as an orthonormal basis for each $1\leq i\leq n$. Then the tensor product of $\mathcal{H}_i$ $(1\leq i\leq n)$ is the Hilbert space with $\{|\psi_{1j_1}...\psi_{nj_n}\rangle\}$ as an orthonormal basis, i.e. \begin{equation*}\begin{split}\bigotimes_{i=1}^{n}\mathcal{H}_i
=\{\sum_{j_1,...,j_n}&\alpha_{j_1...j_n}|\psi_{1j_1}...\psi_{nj_n}\rangle:\\ &\alpha_{j_1...j_n}\in\mathbb{C}\ {\rm for\ all}\ j_1,...,j_n\}
\end{split}\end{equation*} where $|\psi_{1j_1}...\psi_{nj_n}\rangle=|\psi_{1j_1}\rangle ...|\psi_{nj_n}\rangle$ is the product of basis states $|\psi_{1j_1}, ..., |\psi_{nj_n}\rangle$ of the subsystems. In particular, if $\mathcal{H}_i=\mathcal{H}$ for all $1\leq i\leq n$, then $\bigotimes_{i=1}^{n}\mathcal{H}_i$ will be abbreviated to $\mathcal{H}^{\otimes n}$. 

\begin{exam}\label{2ht}The state space of two-qubits is $\mathcal{H}_2^{\otimes 2}$, and a two-qubit system can be in a separable state like $|00\rangle, |1+\rangle$, and it can also be in an entangled state like the EPR pair $$|\beta_{00}\rangle=\frac{1}{\sqrt{2}}(|00\rangle+|11\rangle).$$\hfill $\blacksquare$\end{exam}

\subsection{Density Operators}     
We also assume the reader is familiar with the notion of linear operator. If $\{|i\rangle\}_{i=0}^{n-1}$ is a (fixed) orthonormal basis of an $n-$dimensional Hilbert space $\mathcal{H}$, then an operator $A$ on it can be represented by $n\times n$ matrix $A=(A_{ij})$ where the entries $A_{ij}$ is defined by $$A|i\rangle=\sum_{j=0}^{n-1}A_{ji}|j\rangle$$ for every $0\leq i<n$. An operator $A$ on $\mathcal{H}$ is said
to be positive if $\langle \psi|A|\psi\rangle\geq 0$ for all states
$|\psi\rangle\in \mathcal{H}$. The trace of an operator $A$
is defined to be $$tr(A)=\sum_{i}\langle \psi_i|A|\psi_i\rangle,$$
where $\{|\psi_i\rangle\}$ is an orthonormal basis of $\mathcal{H}$. If the operator is represented by an $n\times n$ matrix $A=\left(A_{ij}\right)$, then its trace is the sum of the entries on the diagonal of $A$, i.e. $tr(A)=\sum_{i=1}^n A_{ii}.$
A mixed state of a quantum system can be described as a density operator when it is not completely known. Let $\{|\psi_i\rangle\}$ be a family of states in $\mathcal{H}$. If a system is in state $|\psi_i\rangle$ with probability $p_i$ for each $i$, and $\sum_ip_i=1$, then the state of the system is represented by $$\rho=\sum_ip_i|\psi_i\rangle\langle\psi_i|,$$ where $|\psi_i\rangle\langle\psi_i|$ is an operator defined as follows: $(|\psi_i\rangle\langle\psi_i|)|\varphi\rangle=\langle\psi_i|\varphi\rangle|\psi_i\rangle$ for each $|\varphi\rangle\in\mathcal{H}$. We say that $\rho$ is a mixed state generated by the ensemble $\{(p_i,|\psi_i\rangle)\}$ of pure states. 
A density operator $\rho$ on a Hilbert space $\mathcal{H}$ is
defined to be a positive operator with $tr(\rho)=1$. An operator is a density operator if and only if it can be generated by an ensemble of pure states.  
In particular, we identify a pure state $|\psi\rangle$ with the density operator $|\psi\rangle\langle\psi|$. 

\begin{exam}The mixed state of a qubit generated by ensemble $\{(\frac{2}{3},|0\rangle),(\frac{1}{3},|1\rangle\}$ is represented by density operator \begin{equation}\label{ex-mix}\rho=\frac{2}{3}|0\rangle\langle 0|+\frac{1}{3}|-\rangle\langle -|=\frac{1}{6}\left (\begin{array}{cc}5 & -1\\ -1&1\end{array}\right )\end{equation}\hfill$\blacksquare$\end{exam}

\subsection{Unitary Operators}
For an operator $A$ on $\mathcal{H}$, if another 
operator $A^{\dag}$ satisfies $(|\varphi\rangle,
A|\psi\rangle)=(A^{\dag}|\varphi\rangle,|\psi\rangle)$ for all
$|\varphi\rangle, |\psi\rangle$, then $A^{\dag}$ is
called the adjoint of $A$, where $(|\chi\rangle,|\zeta\rangle)$ stands for the inner produce $\langle\chi|\zeta\rangle$. An operator $U$ is called a unitary operator 
if $U^{\dag}U=I_{\mathcal{H}}$, where and in the sequel $I_\mathcal{H}$ stands for the identity operator on $\mathcal{H}$. The basic postulate of quantum
mechanics about evolution of systems may be stated as follows:
Suppose that the states of a closed quantum system at times $t_0$
and $t$ are $|\psi_0\rangle$ and $|\psi\rangle$, respectively. Then
they are related to each other by a unitary operator $U$ which
depends only on the times $t_0$ and $t$:
$$|\psi\rangle=U|\psi_0\rangle.$$ This postulate can be reformulated
in the language of density operators as follows. The state $\rho$ of
a closed quantum system at time $t$ is related to its state $\rho_0$
at time $t_0$ by a unitary operator $U$ which depends only on the
times $t$ and $t_0$: $$\rho=U\rho_0 U^{\dag}.$$ A unitary transformation of a state in a finite-dimensional Hilbert space can be calculated by matrix multiplication.

\begin{exam}\label{en-g}  An example of unitary operator on one qubit is the rotation about $x-$axis of the Bloch sphere (see \cite{NC00}, page 19):  
\begin{equation*}R_x(\theta)=\left(\begin{array}{cc} \cos\frac{\theta}{2} & -i\sin\frac{\theta}{2} \\ -i\sin\frac{\theta}{2} & \cos\frac{\theta}{2}\end{array}\right)\end{equation*}
where $0\leq\theta<2\pi$. It transforms the basis state $|0\rangle$ into a superposition of $|0\rangle$ and $|1\rangle$:
\begin{equation*}\begin{split}R_x(\theta)|0\rangle&=\left(\begin{array}{cc} \cos\frac{\theta}{2} & -i\sin\frac{\theta}{2} \\ -i\sin\frac{\theta}{2} & \cos\frac{\theta}{2}\end{array}\right)\left(\begin{array}{cc}1\\0\end{array}\right)\\ &=\left(\begin{array}{cc}\cos\frac{\theta}{2}\\-i\sin\frac{\theta}{2}\end{array}\right)=\cos\frac{\theta}{2}|0\rangle-i\sin\frac{\theta}{2}|1\rangle.\end{split}\end{equation*} 
The controlled-NOT is a unitary operator on two qubits: 
$$CNOT=\left(\begin{array}{cc}I& 0\\ 0&X\end{array}\right),$$ where $I, 0$ are $2\times 2$ unit and zero matrices, respectively, and $$X=\left(\begin{array}{cc}0&1\\1&0\end{array}\right)$$ is the NOT gate. 
The CNOT gate can produce entanglement: $$CNOT(|+0\rangle)=|\beta_{00}\rangle,$$ meaning that separable state $|+0\rangle=|+\rangle|0\rangle$ is transformed to EPR pair $|\beta_{00}\rangle$
\hfill $\blacksquare$ \end{exam}

\subsection{Super-Operators}

A quantum computing or communication system is often not a closed system because it may suffer from unwanted interactions from the environment. 
The dynamics of an open quantum system cannot be described by a unitary
operator, and one of its mathematical formalisms is the notion of
super-operator. A super-operator on a Hilbert space $\mathcal{H}$ is
a linear operator $\mathcal{E}$ from the space of linear operators
on $\mathcal{H}$ into itself which satisfies the following two
conditions: \begin{enumerate} \item $tr[\mathcal{E}(\rho)]\leq
1$ for each density operator $\rho$; \item
Complete positivity: for any extra Hilbert space $\mathcal{H}_R$,
$(\mathcal{I}_R\otimes \mathcal{E})(A)$ is positive provided $A$ is
a positive operator on $\mathcal{H}_R\otimes \mathcal{H}$, where
$\mathcal{I}_R$ is the identity operation on
$\mathcal{H}_R$.\end{enumerate} If condition 1) is strengthened to
$tr[\mathcal{E}(\rho)]=1$ for all density operators $\rho$, then $\mathcal{E}$ is said to be
trace-preserving. In this paper, we only consider trace-preserving super-operators. For any unitary operator $U$, if we define $\mathcal{E}(\rho)=U\rho U^\dag$ for all $\rho$, then $U$ can be seen as a special super-operator $\mathcal{E}$.   

\begin{exam} The \textit{bit flip channel} is widely used in quantum
communication. This channel flips the state of a qubit from
$|0\rangle$ to $|1\rangle$ and vice versa, with probability $1-p$,
$0\leq p\leq 1$. It is described by
the super-operator $\mathcal{E}$ on the $2-$dimensional Hilbert
space $H_2$, defined as follows: $$\mathcal{E}(\rho)=E_0\rho
E_0+E_1\rho E_1$$ for all density operator $\rho$, where
$E_0=\sqrt{p}I,$ $E_1=\sqrt{1-p}X,$ and $I, X$ are the $2\times 2$ unit matrix and the NOT gate, respectively. For example, if $\rho$ is given by Eq.~(\ref{ex-mix}), then it is transformed by $\mathcal{E}$ to another density operator $$\mathcal{E}(\rho)=\left(\begin{array}{cc}\frac{1}{6}+\frac{2p}{3} & -\frac{1}{6}\\ -\frac{1}{6} & \frac{5}{6}-\frac{2p}{3}.
\end{array}\right)$$ \hfill$\blacksquare$ \end{exam}

\subsection{Quantum Measurements}\label{ord-m}
To acquire information about a quantum system, a measurement must be performed on it. In quantum computing, measurement is usually used to read out a computational result. A quantum measurement on a system with state space $\mathcal{H}$ is
described by a collection $\{M_\lambda\}$ of operators
satisfying $$\sum_{\lambda}M^{\dag}_\lambda M_\lambda=I_{\mathcal{H}},$$ where $M_\lambda$ are
called measurement operators, and the indices $\lambda$ stand for the
measurement outcomes. If the state of a quantum system is $|\psi\rangle$ immediately
before the measurement, then the probability that result $\lambda$ occurs
is $$p(\lambda)=\langle \psi|M_\lambda^{\dag}M_\lambda |\psi\rangle$$ and the state of
the system after the measurement is
$$|\psi_\lambda\rangle=\frac{M_\lambda |\psi\rangle}{\sqrt{p(\lambda)}}.$$ We can also
formulate the quantum measurement postulate in the language of
density operators. If the state of a quantum system was $\rho$
immediately before measurement $\{M_\lambda\}$ is performed on it, then
the probability that result $\lambda$ occur is
$$p(\lambda)=tr(M_\lambda^{\dag}M_\lambda\rho),$$ and the state of the system after the
measurement is $$\rho_\lambda=\frac{M_\lambda\rho M_\lambda^{\dag}}{p(\lambda)}.$$

\begin{exam}\label{met}The measurement on a qubit in the computational basis $\{|0\rangle, |1\rangle\}$ is $M=\{M_0,M_1\}$, where $$M_0=|0\rangle\langle 0|=\left(\begin{array}{cc}1&0\\0&0\end{array}\right),\ M_1=|1\rangle\langle 1|=\left(\begin{array}{cc}0&0\\0&1\end{array}\right)$$If we perform $M$ on a qubit in (mixed) state $\rho$ given in Eq.~(\ref{ex-mix}), then the probability that we get outcome $0$ is $$p(0)=tr(M_0\rho)=tr\left(\begin{array}{cc}\frac{5}{6}&0\\0&0\end{array}\right)=\frac{5}{6}$$ and the probability of outcome $1$ is $p(1)=\frac{1}{6}$.
In the case that the outcome is $0$, the qubit will be in state $|0\rangle$ after the measurement, and in the case that the outcome is $1$, it will be in state $|1\rangle$. \hfill$\blacksquare$
\end{exam}

\subsection{POVM Measurements}

In defining noninterference, agents observe the system only at the end, and thus the post-measurement state of the system is of little interest. The Positive-Operator Valued Measure (POVM for short) formalism is especially suited to the analysis of noninterference. A POVM measurement on Hilbert space $\mathcal{H}$ consists of a family of positive operators $\{E_\lambda\}$ such that $$\sum_{\lambda}E_\lambda=I_\mathcal{H}.$$ If it is performed on a system in pure state $|\psi\rangle$, then the probability of outcome $\lambda$ is $$p(\lambda)=\langle\psi|E_\lambda|\psi\rangle;$$ and if the system is in mixed state $\rho$ before measurement, then the probability of outcome $\lambda$ is $$p(\lambda)=tr(E_\lambda\rho).$$ Each ordinary quantum measurement $\{M_\lambda\}$ defined in Subsec.~\ref{ord-m} can be seen as a special POVM measurement if we put $E_\lambda=M^\dag_\lambda M_\lambda$ for all $\lambda$.  

\begin{exam}Let $$E_1=\frac{\sqrt{2}}{1+\sqrt{2}}|1\rangle\langle 1|,\ \ \ \ E_2=\frac{\sqrt{2}}{1+\sqrt{2}}|-\rangle\langle -|$$ and $E_3=I-E_1-E_2,$ where $I$ is the identity operator on the $2-$dimensional Hilbert space. Then $\{E_1,E_2,E_3\}$ is a POVM measurement. If we perform it on a qubit in the state $\rho$ given in Eq.~(\ref{ex-mix}), then the probabilities of outcomes 1, 2 and 3 are, respectively, \begin{equation*}p(1)=\frac{\sqrt{2}}{6(1+\sqrt{2})},\ p(2)=\frac{\sqrt{2}}{3(1+\sqrt{2})},\ p(3)=\frac{2+\sqrt{2}}{2(1+\sqrt{2})}.
\end{equation*}\hfill$\blacksquare$
\end{exam}

\section{Noninterference in Quantum Systems}\label{MOD}

\subsection{An Automata Model of Quantum Systems}

Following Goguen and Meseguer's original formulation~\cite{GM82}, the system models used in the studies of noninterference have been mainly automata. A probabilistic automata model was employed by Gray~\cite{Gr90} in his work on probabilistic (non)interference. Here, we introduce an automata model for quantum systems.    

\begin{defn}A quantum system is a $6-$tuple $$\mathbb{S}=\langle\mathcal{H},\rho_0,A,C,do,measure\rangle,$$ where:\begin{enumerate}\item $\mathcal{H}$ is a Hilbert space, and it is the state space of the system;\item $\rho_0$ is a density operator in $\mathcal{H}$, and it is the initial state;\item $A$ is a set of agents;\item $C$ is a set of commands;  \item $do=\{\mathcal{E}_{a,c}|a\in A\ {\rm and}\ c\in C\}$, and for each $a\in A$ and for each $c\in C$, $\mathcal{E}_{a,c}$ is a super-operator on $\mathcal{H}$, specifying how states are updated by agent $a$ executing command $c$;\item $measure=\{\mathbb{M}_a|a\in A\}$, and for each $a\in A$, $\mathbb{M}_a$ is a set of POVM measurements on $\mathcal{H}$, and intuitively, $\mathbb{M}_a$ consists of all POVM measurements that agent $a$ is allowed to perform.
\end{enumerate}
\end{defn}

The above automata model is defined in a way much more general than that in the majority of quantum automata literature, for example~\cite{MC00}, where only pure states, unitary operators and ordinary (even projective) quantum measurements are considered. Here, we work with the language of density operators (mixed states), super-operators are employed to specify the executions of commands, and POVM measurements are used to describe agents' observation. The major motivation for such a general model is that density operators, super-operators and POVM measurements are commonly adopted in quantum information theory, see for example~\cite{NC00}, Chapter 12. We hope that our results presented in this paper can be smoothly incorporated with quantum information theory to analyse security of quantum computing and communication systems.        

Several essential differences between classical and quantum systems deserve careful explanations. First, the state space of a classical automaton is usually assumed to be discrete and even finite. In this paper, we only consider finite-dimensional quantum automata. But even so, their state Hilbert spaces are a continuum and thus deem-to-be infinite. Second, in the system models of both classical and probabilistic noninterference, the outcomes of agents' observations are deterministic. However, an observation on a quantum system is always done through a quantum measurement which in principle cannot give a deterministic outcome but only a probability distribution of possible outcomes. In addition, an agent may be allowed to observe the system with different measurements which will give different distributions of outcomes.      

Before going ahead, we need to fix some notations. We write $\Sigma^\ast$ for the set of all finite sequences of elements in $\Sigma$. For any $\alpha=\alpha_1\alpha_2\cdots\alpha_n\in (A\times C)^\ast$, the length of $\alpha$ is $|\alpha|=n$. We write $\alpha (i]$ for the head $\alpha_1\alpha_2\cdots\alpha_i$ of $\alpha$ for every $i\leq n$. Also, we write 
$$\mathcal{E}_{\alpha}=\mathcal{E}_{\alpha_n}\circ\cdots\circ\mathcal{E}_{\alpha_2}\circ\mathcal{E}_{\alpha_1}$$ for the composition of $\mathcal{E}_{\alpha_1},\mathcal{E}_{\alpha_2},\cdots,\mathcal{E}_{\alpha_n}$, that is, $$\mathcal{E}_{\alpha}(\rho)=\mathcal{E}_{\alpha_n}(\cdots(\mathcal{E}_{\alpha_2}(\mathcal{E}_{\alpha_1}(\rho)))\cdots)$$ for every density operator $\rho$ in $\mathcal{H}$. 
 
\subsection{Measurement Distance between Density Operators}

Noninterference is defined through a group of agents' nondiscrimination between the final states of the system with and without another group of agents' actions. In the quantum case, observation outcomes are always represented by the probability distributions determined by the involved measurements. So, we first need a distance to measure the difference between two distributions.  
Let $X$ be a finite or countably infinite set. A probability distribution over $X$ is a mapping $p:X\rightarrow [0,1]$ such that $\sum_{x\in X}p(x)=1.$ For each event $E\subseteq X$, the probability of $E$ is given by $$p(E)=\sum_{x\in E}p(x).$$ For any two probability distributions $p$ and $q$ over $X$, their distance is defined to be $$d(p,q)=\frac{1}{2}\sum_{x\in X}|p(x)-q(x)|.$$ It is easy to see that \begin{equation}\label{dim1}d(p,q)=\max_{E\subseteq X}|p(E)-q(E)|.\end{equation} This equality indicates that the distance does not depends on the cardinality of the sample space $X$. 

The above distance between probability distributions can be naturally generalised to a pseudo distance between density operators through quantum measurements. Let $E=\{E_\lambda|\lambda\in\Lambda\}$ be a POVM measurement on $\mathcal{H}$. Then for any density operator $\rho$ in $\mathcal{H}$, we can define a probability distribution $p_E(\rho)=p_E(\rho,\cdot)$ over the measurement outcomes $\Lambda$ by $$p_E(\rho,\lambda)=tr(E_\lambda\rho)$$ for every $\lambda\in\Lambda$. 

Now we consider a family $\mathbb{M}$ of POVM measurements. 

\begin{defn} The pseudo distance defined by $\mathbb{M}$ is given by $$d_\mathbb{M}(\rho,\sigma)=\sup_{E\in \mathbb{M}}d(p_E(\rho),p_E(\sigma))$$ for all density operators $\rho$ and $\sigma$.\end{defn} 

Intuitively, $d_\mathbb{M}(\rho,\sigma)$ measures the difference between $\rho$ and $\sigma$ that can be detected by POVM measurements in $\mathbb{M}$. 

A distance between density operators widely used in quantum information theory is trace distance. Recall from ~\cite{NC00}, Sec. 9.2 that for any density operators $\rho$ and $\sigma$, their trace distance is defined by $$d(\rho,\sigma)=\frac{1}{2}tr|\rho-\sigma|,$$ where $|A|=\sqrt{A^\dag A}$ is the positive square root of $A^\dag A$ for linear operator $A$. The following theorem establishes a connection between trace distance and distance defined by measurements.  

\begin{thm}\label{d-trd} (\cite{NC00}, Theorem 9.1) \begin{equation}\label{dim2}d(\rho,\sigma)=\sup_E d(p_E(\rho),p_E(\sigma)),\end{equation} where the supremum is over all POVM measurements. In other words, if we take $\mathbb{M}$ to be the set of all POVM measurements, then $d(\rho,\sigma)=d_\mathbb{M}(\rho,\sigma).$ \hfill$\blacksquare$\end{thm}

\subsection{(Non)interference Degree}

To present the definition of (non)interference degree, we need several more notations. If $G\subseteq A$ is a group of agents, $D\subseteq C$ is a set of commands, and $\alpha=\alpha_1\alpha_2\cdots\alpha_n\in (A\times C)^\ast$, then following the literature \cite{Ru92, vdM07} on classical noninterference, we write ${\tt purge}_{G,D}(\alpha)$ for the subsequence of $\alpha$ obtained by deleting those $\alpha_i=(a_i,c_i)$ with $a_i\in G$ and $c_i\in D$; that is, 
$${\tt purge}_{G,D}(\alpha)=\alpha_1^\prime\alpha_2^\prime ...\alpha_n^\prime,$$ where $$\alpha_i^\prime=\begin{cases}\epsilon\ &{\rm if}\ \alpha_i=(a,c)\ {\rm with}\ a\in G\ {\rm and}\ c\in D,\\
\alpha_i\ &{\rm otherwise.}\end{cases}$$
We will simply write ${\tt purge}_G(\cdot)$ for ${\tt purge}_{G,C}(\cdot)$. 
For each agent $a\in A$, we write $d_a=d_{\mathbb{M}_a}$ for the pseudo distance defined by the set $\mathbb{M}_a$ of POVM measurements. 

\begin{defn}Let $G_1,G_2\subseteq A$ be two groups of agents, and let $D\subseteq C$ be a set of commands. Then the degree that agents $G_1$ with commands $D$ interfere agents $G_2$ is \begin{equation}\label{up2}\begin{split}
Int(G_1,D|G_2)=\sup\{d_a(&\mathcal{E}_\alpha(\rho_0),\mathcal{E}_{{\tt purge}_{G_1,D}(\alpha)}(\rho_0))\\ &|\alpha\in (A\times C)^\ast, a\in G_2\}.
\end{split}\end{equation}
\end{defn} 

Intuitively, the larger is $Int(G_1,D|G_2)$, the more agents $G_1$ with commands $D$ interfere with agents $G_2$. In particular, if $G_1$ with $D$ does not interfere with $G_2$, that is, \begin{equation}\label{into}\mathcal{E}_\alpha(\rho_0)=\mathcal{E}_{{\tt purge}_{G_1,D}(\alpha)}(\rho_0)\end{equation} for all $\alpha\in (A\times C)^\ast$ and for all $a\in G_2$, then $Int(G_1,D|G_2)=0.$ Conversely, $Int(G_1,D|G_2)=0$ does not necessarily imply Eq.~(\ref{into}) because $d_a$ may not be a distance but only a pseudo distance. In this case, the difference between $\mathcal{E}_\alpha(\rho_0)$ and $\mathcal{E}_{{\tt purge}_{G_1,D}(\alpha)}(\rho_0)$ cannot be detected by agents in $G_2$ using the quantum measurements allowed for them.   
We will simply write $Int(G_1|G_2)$ for $Int(G_1,C|G_2)$. If $Int(G_1,D|G_2)=0$, then we write $G_1,D:|G_2$. Furthermore, we will simply write $G_1:|G_2$ for $G_1,C:|G_2$.

The following proposition considers a special case where agents have the full capacity of measurements, and it follows immediately from Theorem~\ref{d-trd}.

\begin{prop}If each agent $a\in G_2$ can perform any POVM measurement; that is, $\mathbb{M}_a$ is the set of all POVM measurements on $\mathcal{H}$, then 
\begin{equation*}\begin{split}
Int(G_1,D|G_2)=\sup\{d(\mathcal{E}_\alpha(\rho_0),\ & \mathcal{E}_{{\tt purge}_{G_1,D}(\alpha)}(\rho_0))\\ &|\alpha\in (A\times C)^\ast\}.
\end{split}\end{equation*} \hfill$\blacksquare$
\end{prop}

To illustrate the notion defined above, we give a simple example. 
 
\begin{exam}\label{ex1} We consider a system with two qubits. So its state space is $\mathcal{H}_2^{\otimes 2}$, where $\mathcal{H}_2$ is the $2-$dimensional Hilbert space (see Example \ref{ht}). There are two agents Alice and Bob: $A=\{Alice, Bob\}$. They are allowed to perform the measurement in the computational basis (see Example \ref{met}) on the first and second qubits, respectively: $\mathbb{M}_{Alice}=\{M_1\}, \mathbb{M}_{Bob}=\{M_2\}$, where $M_i$ stands for the computational basis measurement on the $i$th qubit for $i=1,2$. The initial state is assumed to be $|\psi_0\rangle=|00\rangle$. 

\begin{enumerate}\item Isolated Alice and Bob: If there is only one command $R_x(\theta)$: $C=\{R_x(\theta)\}$, and when Alice (resp. Bob) execute $R_x(\theta)$, she (resp. he) rotate the first (resp. second) qubit by an angle $\theta$ about the $x-$axis of the Bloch sphere (see Example \ref{en-g}), then the following claim is obvious:  
\begin{itemize}\item 
\textit{Claim:} $Alice:|Bob$ and $Bob:|Alice$; that is, Alice does not interfere with Bob, and vice versa.\end{itemize}  

\item Adding one-way CNOT: Now we add the CNOT gate (see Example \ref{en-g}) into the command set and put $C=\{R_x(\theta),CNOT\}$. Suppose that when both Alice and Bob executes the command $CNOT$, the controlled-NOT transformation is performed with the first qubit as the control qubit and the second as the target qubit. Then we have:
\begin{itemize}
\item \textit{Claim 1:} $Bob, R_x(\theta):|Alice$; that is, $$Int(Bob,R_x(\theta)|Alice)=0,$$ Bob with rotation about $x-$axis does not interfere with Alice. 
\item \textit{Claim 2:} If $\theta\neq 0,\pi$, then $$Int(Bob,CNOT|Alice)>0;$$ that is, Bob with controlled-NOT interferes with Alice.
\item \textit{Claim 3:} If $\theta>0$, then \begin{equation*}\begin{split}&Int(Alice, R_x(\theta)|Bob)>0,\\ &Int(Alice, CNOT|Bob)>0;\end{split}\end{equation*} that is, Alice with either rotation about $x-$axis or controlled-NOT interferes with Bob.\end{itemize}

To prove Claim 1, we notice that each $\alpha\in (A\times C)^\ast$ is a sequence of the following actions:\begin{itemize}
\item $B_1$: Alice execute $R_x(\theta)$ on the first qubit;\item $B_2$: Bob executes $R_x(\theta)$ on the second qubit;\item $B_3$: Alice or Bob executes $CNOT$ with the first qubit as the control qubit and the second as the target qubit.
\end{itemize}
It is obvious that $B_1$ and $B_2$ commute: $B_1B_2=B_2B_1$. Also, it follows from Eq.~(4.39) in~\cite{NC00} that $B_2$ and $B_3$ commute. Suppose that Bob executes $R_x(\theta)$ in $\alpha$ for $n$ times. Then we can move all $R_x(\theta)$ executed by Bob to the end of $\alpha$ and obtain $$\alpha^\prime={\tt purge}(\alpha)(Bob,R_x(\theta))^n,$$ where ${\tt purge}(\alpha)={\tt purge}_{Bob,R_x(\theta)}(\alpha)$ is obtained by deleting all $R_x(\theta)$ executed by Bob from $\alpha$. We write $|\psi\rangle,$ $|\psi^\prime\rangle$, $|\varphi\rangle$ for the states after the system performs $\alpha$, $\alpha^\prime$, and $\mathtt{purge}(\alpha)$, respectively. Then $|\varphi\rangle$ can be written in the following form: $|\varphi\rangle=|0\rangle|\varphi_0\rangle+|1\rangle|\varphi_1\rangle,$ and it holds that 
$$|\psi\rangle=|\psi^\prime\rangle=|0\rangle R_x(n\theta)|\varphi_0\rangle+|1\rangle R_x(n\theta)|\varphi_1\rangle.$$ Finally, Alice measures the first qubit of $|\varphi\rangle$ and $|\psi\rangle$ in the computational basis, she gets the same probability distribution:\begin{equation*}\begin{split}
p(\psi,0)&=||R_x(n\theta)|\varphi_0\rangle||=|||\varphi_0\rangle||=p(\varphi,0),\\ p(\psi,1)&=||R_x(n\theta)|\varphi_1\rangle||=|||\varphi_1\rangle||=p(\varphi,1).
\end{split}\end{equation*}

Now we consider action sequence $\alpha=(Alice,R_x(\theta))(Alice,CNOT)$ $(Bob,CNOT)(Alice,R_x(\theta))$. The state of the system after $\alpha$ is executed is $|\varphi\rangle=(\cos\theta|0\rangle-i\sin\theta|1\rangle)|0\rangle,$ and the state after $\mathtt{purge}_{Bob,CNOT}(\alpha)$ is executed is \begin{equation*}\begin{split}|\psi\rangle=&|0\rangle[\cos^2(\frac{\theta}{2})|0\rangle-\sin^2(\frac{\theta}{2})|1\rangle]\\ &\ \ \ \ \ \ \ \ \ \ -i\sin\frac{\theta}{2}\cos\frac{\theta}{2}|1\rangle(|0\rangle+|1\rangle).
\end{split}\end{equation*} If Alice measures the first qubit of $|\varphi\rangle$ and $|\psi\rangle$ in the computational basis, then the probability distributions of outcomes are \begin{equation*}\begin{split}
&p(0)=\cos^2\theta,\ \ \ \ \ \ \ \ \ \ \ \ \ \ \ \ p(1)=\sin^2\theta,\\ &q(0)=\cos^4(\frac{\theta}{2})+\sin^4(\frac{\theta}{2}), \ q(1)=2\sin^2(\frac{\theta}{2})\cos^2(\frac{\theta}{2}), 
\end{split}\end{equation*} respectively. This implies Claim 2.

To prove Claim 3, consider action sequence $\alpha=(Alice,R_x(\theta))$ $(Alice,CNOT).$ The state becomes $$|\psi\rangle=\cos\frac{\theta}{2}|00\rangle-i\sin\frac{\theta}{2}|11\rangle$$ after executing $\alpha$, and it does not change after executing ${\tt purge}_{Alice,R_x(\theta)}(\alpha)=(Alice,CNOT)$. If Bob measure the second qubit of $|\psi\rangle$ and $|\psi_0\rangle$ in the computational basis, then the probability distributions are \begin{equation*}\begin{split}&p(0)=\cos^2(\frac{\theta}{2}), \ \ \ p(1)=\sin^2(\frac{\theta}{2}),\\ &p_0(0)=1,\ \ \ \ \ \ \ \ \ \  p_0(1)=0,\end{split}\end{equation*}respectively. So, $$Int(Alice,R_x(\theta)|Bob)\geq d(p,p_0)=\sin^2(\frac{\theta}{2})>0.$$  Similarly, we can prove $Int(Alice,CNOT|Bob)>0.$ 

\item Adding two-way CNOT: Finally, we reverse the direction of the CNOT executed by Bob: Suppose that when Bob executes the command CNOT, the second qubit is used as the control qubit and the first qubit is the target. The direction executed by Alice is unchanged. Then we have: \begin{itemize}\item \textit{Claim:} If $\theta>0$, then 
\begin{equation*}\begin{split}&Int(Bob,R_x(\theta)|Alice)>0,\\ &Int(Bob,CNOT|Alice)>0,\\ &Int(Alice, R_x(\theta)|Bob)>0,\\ &Int(Alice, CNOT)|Bob)>0;\end{split}\end{equation*} that is, Alice always interferes with Bob, and vice versa.\end{itemize} 

The proof of this claim is similar to that of the above Claims 2 and 3. \hfill$\blacksquare$ 
\end{enumerate}\end{exam}

The above example indicates that the CNOT gate may cause information leaking in quantum computing. The reason is that certain entanglement between Alice and Bob is created by the CNOT gate.  

\section{Security Policies}\label{SECT}

Information-flow security policies specify how can information flows from one agent to another. We first recall the formal definition of security policy from the literature \cite{Ru92, vdM07} on classical information-flow security.  

\begin{defn} A policy is a reflexive relation between agents: $\leadsto\ \subseteq A\times A.$\end{defn}

Intuitively, $a\leadsto b$ means that actions of agent $a$ are permitted to interfere with agent $b$ or information is permitted to flow from agent $a$ to agent $b$.

Since security policies about a system are only relevant to the rights of agents but not the physical operations in the system such as evolution and observation, it is reasonable to adopt the same definition of policy for classical and quantum systems. Now we can define the notion of security for quantum systems with respect to a given policy based on noninterference. To do so, we need an additional notation. For any agent $a\in A$, we write $\triangledown a=\{b\in A|b\not\leadsto a\}$ for the set of agents from who information cannot flow to agent $a$.   

\subsection{Unbounded-Time Security}

\begin{defn}\label{un-sec}The security degree of system $\mathbb{S}$ with respect to policy $\leadsto$ is \begin{equation}\label{up3}K(\mathbb{S},\leadsto)=\sup_{a\in A}Int(\triangledown a|a).\end{equation}\end{defn}

Intuitively, $Int(\triangledown a|a)$ is the degree that the agents, from whom the policy specifies that information cannot flow to agent $a$, interfere with $a$. 
$K(\mathbb{S},\leadsto)$ takes the supremum of $Int(\triangledown a|a)$ over all agents $a\in A$, and thus measures the global degree that an agent interfere with another agent although information flow from the former to the latter is not allowed by the policy $\leadsto$. Therefore, $K(\mathbb{S},\leadsto)$ can be understood as the degree that system $\mathbb{S}$ is insecure with respect to policy $\leadsto$. The smaller the value of $K(\mathbb{S},\leadsto)$ is, the securer the system $\mathbb{S}$ is. In particular, if $K(\mathbb{S},\leadsto)=0$, then we say that $S$ is secure with respect to $\leadsto$.

\begin{exam}\label{ex2} We extend Example~\ref{ex1} by adding a new agent Charles, so the agent set is $A=\{Alice, Bob, Charles\}$. Consider the security policy $\leadsto$ defined by $Alice\leadsto Bob\leadsto Charles$. The system is expanded to include the third qubit, and the state space is then $\mathcal{H}_2^{\otimes 3}$. The initial state is $|000\rangle$. Alice, Bob and Charles can perform the measurement in the computational basis on the first, second and third qubit, respectively. 

\begin{enumerate}\item Let the command set is $C=\{R_x(\theta)\}$. The executions of $R_x(\theta)$ by Alice and Bob are the same as in Example~\ref{ex1}, and Charles executes $R_x(\theta)$ on the third qubit. It follows immediately from Example~\ref{ex1} 1) that $K(\mathbb{S},\leadsto)=0$; that is, $\mathbb{S}$ is secure with respect to $\leadsto$. 

\item Let the command set $C=\{R_x(\theta),CNOT\}$. 
When Alice executes $CNOT$, the controlled-NOT transformation is performed with the first qubit as the control qubit and the second as the target, and when Bob executes $CNOT$, the controlled-NOT is performed with the second qubit as the control qubit and the third as the target.
Charles is not allowed to execute $CNOT$, or equivalently, when Charles executes $CNOT$, nothing happens. Then it follows from Example~\ref{ex1} 2) that \begin{equation*}\begin{split}&K(\mathbb{S},\leadsto_1)\geq Int(Bob, Charles)|Alice)\\ &\ \ \geq Int(Bob|Alice)\geq Int(Bob, CNOT|Alice)>0,\end{split}\end{equation*} and the system $\mathbb{S}$ is not secure with respect to policy $\leadsto$ when $\theta\neq 0,\pi$. \hfill$\blacksquare$\end{enumerate}\end{exam}

Unwinding is a powerful proof technique for noninterference security of classical systems. 
We can extend the unwinding technique to quantum systems so that it can be used to estimate a upper bound of $K(\mathbb{S},\leadsto)$. A density operator in $\mathcal{H}$ is said to be reachable in system $\mathbb{S}$ if there exists action sequence $\alpha\in (A\times C)^\ast$ such that $\rho=\mathcal{E}_\alpha(\rho_0)$. Then the first version of unwinding theorem can be stated as follows.

\begin{thm}\label{unw} (Unwinding I) If for each agent $a\in A$, there exists an equivalence relation $\stackrel{a}{\sim}$ on reachable density operators satisfying the following conditions:\begin{itemize}
\item Step consistency: $\rho\stackrel{a}{\sim}\sigma \Rightarrow$ $\mathcal{E}_{b,c}(\rho)\stackrel{a}{\sim}\mathcal{E}_{b,c}(\sigma)$ for all $b\in A$ and $c\in C$;\item Local respect of $\leadsto$: $b\not\leadsto a$ $\Rightarrow$ $\rho\stackrel{a}{\sim}\mathcal{E}_{b,c}(\rho)$ for all $c\in C$, 
\end{itemize}
then we have: \begin{equation}\label{up1}\begin{split}K(\mathbb{S},\leadsto)\leq \sup\{&d_a(\rho,\sigma)|\rho\ {\rm and}\ \sigma\ {\rm are}\\ &{\rm reachable},\ \rho\stackrel{a}{\sim}\sigma,\ {\rm and}\ a\in A\}.\end{split}\end{equation}\hfill$\blacksquare$
\end{thm}

Eq.~(\ref{up1}) gives a upper bound of the insecurity degree $K(\mathbb{S},\leadsto)$ under the conditions of Step consistency and Local respect of $\leadsto$. The reader who is familiar with the classical unwinding technique may wonder that Observation consistence seems missing. Indeed, it is incorporated into the right-hand side of Eq. (\ref{up1}). 
In particular, if the equivalence relations $\stackrel{a}{\sim}$, $a\in A$ satisfy the above conditions of Step consistency, Local respect of $\leadsto$ and 
\begin{itemize}
\item Observation consistency: $\rho\stackrel{a}{\sim}\sigma \Rightarrow d_a(\rho,\sigma)=0$; that is, $p_E(\rho)=p_E(\sigma)$ for all POVM measurements $E$ in $M_a$, 
\end{itemize}
then $\mathbb{S}$ is secure with respect to $\leadsto$.

It is known that unwinding proof technique is complete for classical noninterference security~\cite{Ru92, vdM07}. The next theorem shows that the unwinding proof technique presented in Theorem~\ref{unw} is complete for absolute security of quantum systems. 

\begin{thm}\label{unw1} (Partial Completeness of Unwinding I) If $\mathbb{S}$ is secure with respect to $\leadsto$, then there exists a family $\stackrel{a}{\sim},$ $a\in A$ of equivalence relations on reachable density operators satisfying Step consistency, Observation consistency and Local respect of $\leadsto$. \hfill$\blacksquare$
\end{thm}

\subsection{Bounded-Time Security}

Note that the length of action sequence $\alpha$ in Eq. (\ref{up2}) can be arbitrary. Thus, in the definition equation (\ref{up3}) of insecurity degree $K(\mathbb{S},\leadsto)$, the time used by malicious agents to detect sensitive information is unlimited. We now consider a bounded-time variant of Definition~\ref{un-sec}.

\begin{defn}Let $t$ be a nonnegative real number. Then the degree that system $\mathbb{S}$ is $t-$bounded insecure with respect to policy $\leadsto$ is \begin{equation*}\begin{split}K_t(\mathbb{S},\leadsto)=\sup\{d_a&(\mathcal{E}_\alpha(\rho_0), \mathcal{E}_{{\tt purge}_{\triangledown a}(\alpha)}(\rho_0)\\ &|\alpha\in (A\times C)^\ast, |\alpha|\leq t, a\in A\}.\end{split}\end{equation*}\end{defn}

Intuitively, $K_t(\mathbb{S},\leadsto)$ measure the (in)security degree of system $\mathbb{S}$ with respect to policy $\leadsto$ under the assumption that the running time of the system does not exceeds $t$. It is obvious that $K_t(\mathbb{S},\leadsto)\leq K_{t^\prime}(\mathbb{S},\leadsto)$ if $t\leq t^\prime$. If particular, if $K_t(\mathbb{S},\leadsto)=0$, then we say that $\mathbb{S}$ is secure with respect to $\leadsto$ within time $t$. 

We have a unwinding proof technique for bounded-time security too.

\begin{thm}\label{unw2}(Unwinding II) Let $\epsilon_s, \epsilon_o, \epsilon_l$ be nonnegative real numbers. If for each agent $a\in A$, there exists a pseudo-distance $\delta_a$ between reachable density operators satisfying the following conditions: \begin{itemize}\item Approximate step consistency:
$$\delta_a(\mathcal{E}_{b,c}(\rho),\mathcal{E}_{b,c}(\sigma))\leq \delta_a(\rho,\sigma)+\epsilon_s$$ for all $b\in A$ and $c\in C$, and for all $\rho,\sigma$; \item Approximate observation consistency: 
$$d_a(\rho,\sigma)\leq \delta_a(\rho,\sigma) + \epsilon_o$$ for all $\rho,\sigma$; 
\item Approximate local respect of $\leadsto$:  if $b\not\leadsto a$, then it holds that $$\delta_a(\rho,\mathcal{E}_{b,c}(\rho))\leq\epsilon_l$$ for all $c\in C$, and for all $\rho,\sigma$,    
\end{itemize}then we have: \begin{equation}\label{up4}K_t(\mathbb{S},\leadsto)\leq \epsilon_o+ t\cdot\max \{\epsilon_s,\epsilon_l\}.\end{equation} \hfill$\blacksquare$
\end{thm}

A upper bound of insecurity degree $K_t(\mathbb{S},\leadsto)$ is given by Eq. (\ref{up4}). 
The next theorem derives a lower bound of insecurity degree $K(\mathbb{S},\leadsto)$ through bounded-time security $K_t(\mathbb{S},\leadsto)$.

\begin{thm}\label{unw3}(Weak Completeness of Unwinding II) There exists a family $\delta_a, a\in A$ of pseudo-distance on reachable density operators satisfying the following conditions:
\begin{itemize}\item Step consistency:
$$\delta_a(\mathcal{E}_{b,c}(\rho),\mathcal{E}_{b,c}(\sigma))\leq \delta_a(\rho,\sigma)$$ for all $b\in A$ and $c\in C$, and for all $\rho,\sigma$; \item Observation consistency: 
$$d_a(\rho,\sigma)\leq \delta_a(\rho,\sigma) $$ for all $\rho,\sigma$; 
\item Bounded local respect of $\leadsto$:
\begin{equation}\label{up5}\begin{split}\lim_{t\rightarrow t^\prime} &K_t(\mathbb{S},\leadsto)=K(\mathbb{S},\leadsto)\\ 
&\geq \frac{1}{2}\sup\{\delta_a(\rho,\mathcal{E}_{b,c}(\rho)|b\not\leadsto a, c\in C\\ &\ \ \ \ \ \ \ \ \ \ \ \ \ \ \ \ \ \ \ \ \ \ {\rm and}\ \rho\ {\rm is\ reachable}\}.
\end{split}\end{equation}\end{itemize} \hfill$\blacksquare$ 
\end{thm}

The lower bound of insecurity degree $K(\mathbb{S},\leadsto)$ in Eq. (\ref{up5}) can be seen as a weak completeness of the unwinding technique presented in Theorem \ref{unw2}. 
In particular, if $\mathbb{S}$ is secure with respect to $\leadsto$; that is, $K(\mathbb{S},\leadsto)=0$, then there exists a family $\delta_a$, $a\in A$ of pseudo-distances on reachable density operators satisfying the above Step and Observation consistency and the following:
\begin{itemize}\item Local respect of $\leadsto$: $b\not\leadsto a\Rightarrow \delta_a(\rho,\mathcal{E}_{b,c}(\rho))=0$ for all $c\in A$ and for all $\rho$.\end{itemize} 

\subsection{Strong Security}

Different from classical systems, the state of a quantum system is often not completely known and thus the system is in a mixed state defined by a statistical ensemble. Some stronger security degrees will be useful when we consider mixtures of initial states. 
Let $\rho$ be a density operator, $\{p_i\}$ a probability distribution, and $\rho_i$ a density operator for every $i$. If $\rho=\sum_i p_i\rho_i,$ then $\rho$ is called a mixture of ensemble $\{(p_i,\rho_i)\}$ of density operators.

Before presenting the definition of strong security degree, we have to introduce a notation. Let $\mathbb{S}=\langle\mathcal{H},\rho_0,A,C,do,messure\rangle,$ and let $\rho_0^\prime$ be a density operator. We write $\mathbb{S}[\rho_0^\prime]$ for the new system obtained by replacing the initial state $\rho_0$ in $\mathbb{S}$ by another initial state $\rho_0^\prime$; that is, 
$$\mathbb{S}[\rho_0^\prime]=\langle\mathcal{H},\rho_0^\prime,A,C,do,messure\rangle.$$

\begin{defn}\begin{enumerate}\item The strong security degree of system $\mathbb{S}$ with respect to $\leadsto$ is 
\begin{equation}\label{def-ss}\begin{split}SK&(\mathbb{S},\leadsto)=\sup\{\sum_ip_iK(\mathbb{S}[\rho_i],\leadsto):\\ &\rho_0\ {\rm is\ a\ mixture\ of\ ensemble}\ \{(p_i,\rho_i)\}\}.
\end{split}\end{equation}\item Let $t$ be a positive real number. Then the strong $t-$bounded security degree $SK_t(\mathbb{S},\leadsto)$ of $\mathbb{S}$ with respect to $\leadsto$ is defined by Eq.~(\ref{def-ss}) with $K$ substituted by $K_t$. 
\end{enumerate}
\end{defn}

\section{Compositionality of Security}\label{COMPS}

The purpose of this section is to examine security of a system composed of a collection of subsystems.
We consider two quantum systems $$\mathbb{S}=\langle\mathcal{H},\rho_0, A, C, do, measure\rangle,$$
$$\mathbb{S}^\prime=\langle\mathcal{H}^\prime,\rho_0^\prime, A^\prime, C^\prime, do^\prime, measure^\prime\rangle.$$ We can assume that $C\cap C^\prime=\emptyset$ without any loss of generality because the commands in $C$ are executed on the component $\mathbb{S}$, whereas the commands in $C^\prime$ are executed on the different component $\mathbb{S}^\prime$. However, it is allowed that $A\cap A^\prime=\emptyset$ because the same agent may be granted to access both components $\mathbb{S}$ and $\mathbb{S}^\prime$. 

\begin{defn}\label{comp-def1}
The composition of $\mathbb{S}$ and $\mathbb{S}^\prime$ is defined to be the quantum system  $$\mathbb{S}\otimes\mathbb{S}^\prime=\langle\mathcal{H}\otimes\mathcal{H}^\prime,\rho_0\otimes\rho_0^\prime,A\cup A^\prime, C\cup C^\prime, Do, Measure\rangle,$$ where \begin{enumerate}
\item $Do=\{\mathcal{F}_{a,c}|a\in A\cup A^\prime\ {\rm and}\ c\in C\cup C^\prime\}$, \begin{equation}\label{comp1}
\mathcal{F}_{a,c}=\begin{cases}\mathcal{E}_{a,c}\otimes\mathcal{I}_{\mathcal{H}^\prime}\ &{\rm if}\ a\in A\ {\rm and}\ c\in C,\\ 
\mathcal{I}_{\mathcal{H}}\otimes\mathcal{E}_{a,c}^{\prime}\ &{\rm if}\ a\in A^\prime\ {\rm and}\ c\in C^\prime,\\
\mathcal{I}_{\mathcal{H}\otimes\mathcal{H}^\prime}\ &{\rm if}\ a\in A\setminus A^\prime\ {\rm and}\ c\in C^\prime,\\ & {\rm or}\ a\in A^\prime\setminus A\ {\rm and}\ c\in C; 
\end{cases}
\end{equation}\item $Measure=\{\mathbb{N}_a|a\in A\cup A^\prime\}$, \begin{equation}\label{comp2}
\mathbb{N}_a=\begin{cases}\mathbb{M}_a\ &{\rm if}\ a\in A\setminus A^\prime,\\ \mathbb{M}_a\cup\mathbb{M}_a^\prime\ &{\rm if}\ a\in A\cap A^\prime,\\ \mathbb{M}_a\ &{\rm if}\ a\in A^{\prime}\setminus A.
\end{cases}
\end{equation}
\end{enumerate}
\end{defn}

To simplify presentation, a little bit of notation abuse was allowed in the defining equation of $\mathbb{N}_a$; for example, if $E\in\mathbb{M}_a$ and $a\in A$, then $E$ is a measurement on the whole system when it is considered in $\mathbb{S}$, but it is a measurement on a subsystem $\mathbb{S}$ when it is considered in $\mathbb{S}\otimes\mathbb{S}^\prime$. 

We also consider the combination of two security policies. To this end, we need a notation.  
Let $R\subseteq X\times X$ be a binary relation on $X$, and let $Y\subseteq X$. Then we write $R|Y$ for the restriction of $R$ on $Y$; that is, $R|Y=\{(x,y)\in Y\times Y:xRy\}.$

\begin{defn}Let $\leadsto$ be a policy for agents $A$ and $\leadsto^\prime$ a policy for agents $A^\prime$. \begin{enumerate}\item If $\leadsto |A\cap A^\prime=\leadsto^\prime |A\cap A^\prime,$ then we say that $\leadsto$ and $\leadsto^\prime$ are compatible. \item The union of $\leadsto$ and $\leadsto^\prime$ is the policy $\leadsto\cup\leadsto^\prime$ on agents $A\cup A^\prime$. 
\end{enumerate}
\end{defn}

Now we are ready to prove that security of quantum systems is compositional.  

\begin{thm}\label{comp-sec0}If $\leadsto$ and $\leadsto^\prime$ are compatible, then we have: \begin{enumerate}\item $K(\mathbb{S}\otimes\mathbb{S}^\prime,\leadsto\cup\leadsto^\prime)\leq K(\mathbb{S},\leadsto)+K(\mathbb{S}^\prime,\leadsto^\prime);$
\item $K_t(\mathbb{S}\otimes\mathbb{S}^\prime,\leadsto\cup\leadsto^\prime)\leq K_t(\mathbb{S},\leadsto)+K_t(\mathbb{S}^\prime,\leadsto^\prime).$ \hfill$\blacksquare$
\end{enumerate}\end{thm}

The above theorem shows that the insecurity degree of a composed system does not exceed the sum of the insecurity degrees of its component systems. In particular, if $\mathbb{S}$ and $\mathbb{S}^\prime$ are secure with respect to $\leadsto$ and $\leadsto^\prime$, respectively (within time $t$), then $\mathbb{S}\otimes\mathbb{S}^\prime$ is secure with respect to $\leadsto\cup\leadsto^\prime$ (within time $t$).

The composition of quantum systems in Definition~\ref{comp-def1} is indeed a direct product in which the component systems are entirely independent to each other except that some agent can access to different components. We can introduce a more general notion of composition where component quantum systems can be hooked up more tightly. To define it, we need several auxiliary notions. 
Recall from~\cite{NC00} that the partial trace $tr_{\mathcal{H}^\prime}$ over $\mathcal{H}^\prime$ is a mapping from density operators in $\mathcal{H}\otimes\mathcal{H}^\prime$ to density operators in $\mathcal{H}$. It is defined by $$tr_{\mathcal{H}^\prime}(|\varphi_1\rangle\langle\varphi_2|\otimes|\psi_1\rangle\langle\psi_2|)=\langle\psi_2|\psi_1\rangle|\varphi_1\rangle\langle\varphi_2|$$ for all $|\varphi_1\rangle,|\varphi_2\rangle\in\mathcal{H}_1$ and $|\psi_1\rangle,|\psi_2\rangle\in\mathcal{H}^\prime$, and it is extended to all density operators in $\mathcal{H}\otimes\mathcal{H}^\prime$ by linearity. 
Let $\rho$ be a density operator in $\mathcal{H}$ and $\sigma$ a density operator in $\mathcal{H}\otimes\mathcal{H}^\prime$. 
If $tr_{\mathcal{H}^\prime}(\sigma)=\rho$, then $\sigma$ is called an extension of $\rho$ in $\mathcal{H}\otimes\mathcal{H}^\prime$.  Let $\mathcal{E}$ be a super-operator on $\mathcal{H}$ and $\mathcal{F}$ a super-operator on $\mathcal{H}\otimes\mathcal{H}^\prime$. We say that $\mathcal{F}$ is a cylindrical extension of $\mathcal{E}$ on in $\mathcal{H}\otimes\mathcal{H}^\prime$ if \begin{enumerate}\item $tr_{\mathcal{H}^\prime}\mathcal{F}(\rho\otimes\frac{1}{d^\prime}I_{\mathcal{H}^\prime})=\mathcal{E}(\rho)$ for all density operators in $\mathcal{H}$; \item $tr_\mathcal{H}\mathcal{F}(\frac{1}{d}I_\mathcal{H}\otimes\rho^\prime)=\rho^\prime$ for all density operators $\rho^\prime$ in $\mathcal{H}^\prime$,\end{enumerate} where $d=\dim\mathcal{H}$ and $d^\prime=\dim\mathcal{H}^\prime$. 

\begin{defn}\label{comp-def2}
A generalised composition of $\mathbb{S}$ and $\mathbb{S}^\prime$ is defined to be a quantum system  $$\mathbb{T}=\langle\mathcal{H}\otimes\mathcal{H}^\prime,\sigma_0,A\cup A^\prime, C\cup C^\prime, Do, Measure\rangle,$$ where \begin{enumerate}\item $\sigma_0$ is both an extension of $\rho_0$ and an extension of $\rho_0^\prime$ in 
$\mathcal{H}\otimes\mathcal{H}^\prime$; \item $Do=\{\mathcal{F}_{a,c}|a\in A\cup A^\prime\ {\rm and}\ c\in C\cup C^\prime\}$ satisfies the following conditions:\begin{itemize}
\item $\mathcal{F}_{a,c}=\mathcal{I}_{\mathcal{H}\otimes\mathcal{H}^\prime}$ if $a\in A\setminus A^\prime$ and $c\in C^\prime$, or $a\in A^\prime\setminus A$ and $c\in C;$ \item $\mathcal{F}_{a,c}$ is a cylindrical extension of $\mathcal{E}_{a,c}$ if $a\in A$ and $c\in C$; \item $\mathcal{F}_{a,c}$ is a cylindrical extension of $\mathcal{E}^\prime_{a,c}$ if $a\in A^\prime$ and $c\in C^\prime$;
\end{itemize} \item $Measure$ is the same as in Definition~\ref{comp-def1}.\end{enumerate}\end{defn}

Theorem~\ref{comp-sec0} can be extended to a special class of generalised compositions of quantum systems. Recall from~\cite{Ho09} that a density operator $\sigma$ in $\mathcal{H}\otimes\mathcal{H}^\prime$ is said to be separable if we can write: $$\sigma=\sum_ip_i\rho_i\otimes\rho_i^\prime$$ where all $\rho_i$ are density operators in $\mathcal{H}$ and all $\rho_i^\prime$ in $\mathcal{H}^\prime$, $p_i\geq 0$ for all $i$, and $\sum_ip_i=1$. A super-operator $\mathcal{F}$ on $\mathcal{H}\otimes\mathcal{H}^\prime$ is said to be separable if there are a family $\{F_i\}$ of operators on $\mathcal{H}$ and a family $\{F_i^\prime\}$ of operators on $\mathcal{H}^\prime$ such that $$\mathcal{F}(\sigma)=\sum_i(F_i\otimes F_i^\prime)\sigma(F_i^\dag\otimes F_i^{\prime\dag})$$ for all density operators $\sigma$ on $\mathcal{H}\otimes\mathcal{H}^\prime$. If operators $F_i$ commute, i.e. $F_iF_j=F_jF_i$ for all $i\neq j$, and operators $F_i^\prime$ commute, then $\mathcal{F}$ is said to be commutative. 

\begin{thm}\label{Comp-SEC}If $\leadsto$ and $\leadsto^\prime$ are compatible, and $\mathbb{T}$ is a generalised composition of $\mathbb{S}$ and $\mathbb{S}^\prime$ with a separable initial state $\sigma_0$ and commutative and separable super-operators $\mathcal{F}_{a,c}$ ($a\in A\cup A^\prime$, $c\in C\cup C^\prime$), then we have: \begin{enumerate}\item $K(\mathbb{T},\leadsto\cup\leadsto^\prime)\leq SK(\mathbb{S},\leadsto)+SK(\mathbb{S}^\prime,\leadsto^\prime);$
\item $K_t(\mathbb{T},\leadsto\cup\leadsto^\prime)\leq SK_t(\mathbb{S},\leadsto)+SK_t(\mathbb{S}^\prime,\leadsto^\prime).$ \hfill$\blacksquare$
\end{enumerate}
\end{thm}

\section{Access Control}\label{ACCC}

As an application of the quantum noninterference formalism developed in the previous sections, we now analyse security of access control to quantum information. To do so, we impose certain internal structure on the system under consideration by assuming that information is stored in different locations. 

\begin{defn}We say that the system $\mathbb{S}$ has structured states if there exists a set $N$ of location names, and for each location name $n\in N$, there exists a Hilbert space $\mathcal{H}_n$ such that $$\mathcal{H}=\bigotimes_{n\in N}\mathcal{H}_n.$$
\end{defn}

In other words, the quantum system $\mathbb{S}$ is a composed system that consists of component systems labeled by locations $n\in N$. 

There is an essential difference between quantum and classical systems that makes understanding access control in a quantum system harder than that in a classical system. In a classical system, access control is usually defined by a matrix consisting of two functions \textquotedblleft $read$" and \textquotedblleft $alter$", specifying whether a given agent may \textquotedblleft read", \textquotedblleft alter", respectively the information stored in given locations; for example, for each $a\in A$, $read(a)$ is defined to be a subset of location names $N$, and it is the set of locations whose values can be read by agent $a$. The reasonableness of defining $read(a)$ as a subset of $N$ comes from an implicit assumption:  
\begin{itemize}\item \textquotedblleft$1+1= 1$": The ability to observe both the $K$ subsystem (i.e. the subsystem consisting components labeled by $n\in K$) and the $L$ subsystem implies the ability to observe the combined $K\cup L$ subsystem, where $K,L\subseteq N$.
\end{itemize} Whenever this assumption is not valid, then $read(a)$ must be defined as a subset of $\mathcal{P}(N)$ instead of a subset of $N$, where we use $\mathcal{P}(\cdot)$ to denote power set; for example, suppose that $N=\{n_1,n_2,n_3\}$. If agent $a$ is allowed to read both the values of location $n_1$ and $n_2$ but not the value of combined location $n_1n_2$, then $read(a)=\{\{n_1\},\{n_2\}\}$; if agent $a$ is allowed to read the values of location $n_1$ and $n_2$ as well as $n_1n_2$, then $read(a)=\{\{n_1\},\{n_2\},\{n_1,n_2\}\}$. Indeed, the above \textquotedblleft$1+1=1$" assumption is violated in the quantum world, as indicated by the next example. For simplicity, for any $K\subseteq N$, we write $tr_K$ for the partial trace $tr_{\bigotimes_{n\in K}\mathcal{H}_n}$ over the $K$ subsystem. 
\begin{exam}\label{112} There are $\rho,\sigma\in\mathcal{H}=\bigotimes_{n\in N}\mathcal{H}_n$ such that \begin{enumerate}\item $tr_{N\setminus K}(\rho)=tr_{N\setminus K}(\sigma)$ and $tr_{N\setminus L}(\rho)=tr_{N\setminus L}(\sigma)$; but \item $tr_{N\setminus (K\cup L)}(\rho)\neq tr_{N\setminus (K\cup L)}(\sigma)$.\end{enumerate} 
In this case, an agent who can read information stored in $K$ and information stored in $L$ but not information stored in $K\cup L$ is unable to distinguish $\rho$ from $\sigma$.  
For instance, let $N=\{n_1,n_2\}$, and let $\mathcal{H}_{n_1}=\mathcal{H}_{n_2}$ be the $2-$dimensional Hilbert space $\mathcal{H}_2$. We put $$\rho=\frac{1}{2}|00\rangle\langle 00|+\frac{1}{2}|11\rangle\langle 11|,$$ and $\sigma=|\beta_{00}\rangle\langle\beta_{00}|$, where $|\beta_{00}\rangle=\frac{1}{\sqrt{2}}(|00\rangle+|11\rangle)$ is the EPR pair. 
Then \begin{equation*}\begin{split}
tr_{n_1}(\rho)&=tr_{n_1}(\sigma)=\frac{1}{2}(|0\rangle\langle 0|+|1\rangle\langle 1|,\\
tr_{n_2}(\rho)&=tr_{n_2}(\sigma)=\frac{1}{2}(|0\rangle\langle 0|+|1\rangle\langle 1|,
\end{split}\end{equation*} but it is obvious that $\rho\neq\sigma$. \hfill$\blacksquare$
\end{exam}  

Similarly, in the quantum world we know: 
\begin{itemize}
\item The ability to change both the state of the $K$ subsystem and the state of the $L$ subsystem does not guarantee the ability to change the combined $K\cup L$ subsystem.
\end{itemize}
\begin{exam}There are $\rho\in\mathcal{H}_{N\setminus K}$ and $\sigma\in\mathcal{H}_{N\setminus L}$ such that \begin{enumerate}\item $tr_{N\setminus (K\cap L)}(\rho)=tr_{N\setminus (K\cap L)}(\sigma)$; but \item there does not exist ant $\gamma\in\mathcal{H}_{N\setminus (K\cup L)}$ such that $tr_{N\setminus K}(\gamma)=\rho$ and $tr_{N\setminus L}(\gamma)=\sigma$. \hfill$\blacksquare$
\end{enumerate}
\end{exam}
Another essential difference between quantum and classical information is that reading the quantum information stored in a certain location changes the information itself; but this difference will not be considered in this paper because in the noninterference formalism reading (by quantum measurements) always happens at the end, and thus the post-measurement state of the system is irrelevant.   

By the above observation, we realise that both $read(a)$ and $alter(a)$ should be defined as elements of $\mathcal{P}(\mathcal{P}(N))$. They can be simplified a little bit by noticing that if an agent can read (resp. alter) the value of locations $K$ then it can read (resp. alter) the value of any subset $L$ of $K$. 
A family $\mathcal{B}\in\mathcal{P}(\mathcal{P}(N))$ of sets of location names is said to be below-closed if $$K\in\mathcal{B}\ {\rm and}\ L\subseteq K\Rightarrow L\in\mathcal{B}.$$ We write $\mathcal{P}_B(\mathcal{P}(N))$ for the set of all below-closed $\mathcal{B}\in\mathcal{P}(\mathcal{P}(N))$. 

\begin{defn}An access control matrix consists of:\begin{enumerate}\item a function $read:A\rightarrow\mathcal{P}_B(\mathcal{P}(N))$; and \item a function $alter:A\rightarrow\mathcal{P}_B(\mathcal{P}(N))$, 
\end{enumerate}
\end{defn}

For each agent $a\in A$, if $K\in read(a)$, then the $K$ subsystem can be observed by $a$; and if $K\in alter(a)$, then the state of the $K$ subsystem can be changed by $a$.   

We now consider security of quantum access control with respect to a policy.   

\begin{defn}\label{acm-s}An access control matrix $(read,alter)$ satisfies security policy $\leadsto$ if \begin{enumerate}\item $a\leadsto b\Rightarrow read(a)\subseteq read(b)$; \item $(\exists K\in read(a), \exists L\in alter(b)$ s.t. $K\cap L\neq\emptyset)\Rightarrow b\leadsto a$.
\end{enumerate}
\end{defn}

To present the quantum generalisation of Rushby's security theorem for access control~\cite{Ru92}, we need to introduce a new pseudo-distance between density operators. For each agent $a\in A$, we define distance $\delta_a$ by $$\delta_a(\rho,\sigma)=\sup_{K\in read(a)} d(tr_{N\setminus K}(\rho),tr_{N\setminus K}(\sigma))$$ for all reachable density operators $\rho,\sigma$ in $\mathcal{H}=\bigotimes_{n\in N}\mathcal{H}_n$. Intuitively, $\delta_a(\rho,\sigma)$ measures the difference between $\rho$ and $\sigma$ at the locations that agent $a$ can observe. Note that in the defining equation of $\delta_a$, the supremum is taken over the distances $d$ in some subspaces of $\mathcal{H}$ of different dimensions. From Eqs.~(\ref{dim1}) and~(\ref{dim2}) we see that the distances $d$ does not depends on the dimensions of these subspaces, so this defining equation is not problematic.     

Let $\epsilon>0$ and $K\subseteq N$. For any density operators $\rho,\sigma\in\mathcal{H}$, if $$d(tr_{N\setminus K}(\rho),tr_{N\setminus K}(\sigma))>\epsilon,$$ then we say that $\rho$ and $\sigma$ are $\epsilon-$discriminable on $K$, and write $Dis(\rho,\sigma|\epsilon, K)$.
Now we are ready to present the main result of this section, which gives a upper bound of bounded-time insecurity degree in terms of Reference Monitor Assumption and thus generalises Theorem 2 of \cite{Ru92} to the quantum case. 

\begin{thm}\label{main-acc}If the access control matrix satisfies policy $\leadsto$ and the Reference Monitor Assumptions: for all $a\in A$, for all $c\in C$, for all $\rho,\sigma$, and for all $K\subseteq N$, 

(RM1) $d_a(\rho,\sigma)\leq\delta_a(\rho,\sigma)+\theta$;

(RM2) \begin{equation*}\begin{split}\ \ \ \ \ \ \ \ \ Dis(\rho,\mathcal{E}_{a,c}&(\rho)|\epsilon, K)\vee Dis(\sigma, \mathcal{E}_{a,c}(\sigma)|\epsilon, K)\\ &\Rightarrow \neg Dis(\mathcal{E}_{a,c}(\rho),\mathcal{E}_{a,c}(\sigma)
|\delta_a(\rho,\sigma), K);\end{split}\end{equation*}

(RM3) $$Dis(\rho,\mathcal{E}_{a,c}(\rho)|\epsilon, K)\Rightarrow \exists L\in alter(a)\ {\rm s.t.}\ K\cap L\neq\emptyset,$$
then it holds that $K_t(\mathbb{S},\leadsto)\leq \theta+2t\epsilon.$ \hfill$\blacksquare$
\end{thm}  

\section{Conclusion}\label{concl}
 
 The noninterference formalism of information-flow security is generalised to the quantum case. We define three (in)security degrees $K(\mathbb{S},\leadsto)$, $K_t(\mathbb{S},\leadsto)$ and $SK(\mathbb{S},\leadsto)$ of a quantum system modelled by a quantum automaton $\mathbb{S}$ with respect to a security policy $\leadsto$. The unwinding technique for proving noninterference security is extended so that it can be used to give a upper bound of the (in)security degrees of quantum systems. A compositionality theorem for security of quantum systems is established, showing that the (in)security degree of a composite system does not exceed the sum of the (in)security degrees of its components. 
 
For further research, one open question is to settle the computational complexity of the following problem:  given a quantum system $\mathbb{S}$, a security policy $\leadsto$, and a rational constant $c$, decide whether $K(\mathbb{S},\leadsto)<c$, $K_t(\mathbb{S},\leadsto)<c$, and $SK(\mathbb{S},\leadsto)<c$?

Only transitive noninterference for quantum systems is considered in this paper. As argued in~\cite{HY87},~\cite{Ru92}, transitive policies are too restrictive for many realistic applications, and since then intransitive noninterference for classical systems has been intensively studied; see for example~\cite{Ro99}, \cite{vdM07}, \cite{MZ10}. So, another topic for further research is to define intransitive noninterference for quantum systems. 

Noninterference was also defined by Focardi and Gorrieri~\cite{FG95} and Ryan and Schneider~\cite{RS01} in the framework of process algebras based on the notion of process equivalence.  
Several quantum processes have been defined in the last decade, including Jorrand and Lalire's QPAlg~\cite{JL04}, Gay and Nagarajan's CQP~\cite{GN05} and the authors' qCCS~\cite{FDY11}, \cite{FDY12}. In particular, a bisimilarity preserved by parallel composition of quantum processes with entanglement was recently discovered by the authors~\cite{FDY11, FDY12} and Davidson~\cite{Dav11}. A process equivalence-based quantum interference would be another interesting topic.     
 
\section*{Acknowledgment}
This work was partly supported by the Australian Research Council (Grant No: DP110103473 and FT100100218) and the Overseas Team Program of Academy of Mathematics and Systems Science, Chinese Academy of Sciences.

\newpage

\section*{Appendix: Proofs of Main Results}

\subsection*{A. Proofs of Theorems in Section~\ref{SECT}}

\subsubsection*{A.1. Proof of Theorem~\ref{unw}} 

By definition, we have: 
\begin{equation*}\begin{split}K(\mathbb{S},\leadsto)=\sup\{d_a(&\mathcal{E}_\alpha(\rho_0),\mathcal{E}_{{\tt purge}_{\triangledown a}(\alpha)}(\rho_0))\\ &|\alpha\in (A\times C)^\ast\ {\rm and}\ a\in A\}.\end{split}\end{equation*}So, it suffices to show that for each $a\in A$ and for each $\alpha\in (A\times C)^\ast$, $$\mathcal{E}_\alpha(\rho_0)\stackrel{a}{\sim}\mathcal{E}_{{\tt purge}_{\triangledown a}(\alpha)}(\rho_0).$$This can be easily done by induction on the length of $\alpha$, and we omit the routine details. \hfill$\blacksquare$

\subsubsection*{A.2. Proof of Theorem~\ref{unw1}}

For each agent $a\in A$, we define: $$\rho\stackrel{a}{\sim}\sigma\Leftrightarrow d_a(\mathcal{E}_\alpha(\rho),\mathcal{E}_\alpha(\sigma))=0\ {\rm for\ all}\ \alpha\in (A\times C)^\ast.$$
It is easy to see that $\stackrel{a}{\sim},$ $a\in A$ satisfy Step and Observation consistency. To show that they locally respect $\leadsto$, we assume that $b\not\leadsto a$. Then for any reachable density operator $\rho$, we have $\rho=\mathcal{E}_\beta(\rho_0)$ for some action sequence $\beta\in (A\times C)^\ast$. Furthermore, for any $\alpha\in (A\times C)^\ast$ and for any $c\in C$, it holds that $${\tt purge}_{\triangledown a}(\beta\alpha)={\tt purge}_{\triangledown a}(\beta(b,c)\alpha).$$ Therefore, we have \begin{equation*}\begin{split}
d_a(&\mathcal{E}_\alpha(\rho),\mathcal{E}_\alpha(\mathcal{E}_{b,c}(\rho)))=d_a(\mathcal{E}_{\beta\alpha}(\rho_0),\mathcal{E}_{\beta(b,c)\alpha}(\rho_0))\\
&\leq d_a(\mathcal{E}_{\beta\alpha}(\rho_0),\mathcal{E}_{{\tt purge}_{\triangledown a}(\beta\alpha)}(\rho_0))+\\ &\ \ \ \ \ \ \ \ \ \ \ \ \ \ \ \ \ \ \ \ \ \ \ d_a(\mathcal{E}_{{\tt purge}_{\triangledown a}(\beta\alpha)}(\rho_0),\mathcal{E}_{\beta(b,c)\alpha}(\rho_0))\\ &=d_a(\mathcal{E}_{\beta\alpha}(\rho_0),\mathcal{E}_{{\tt purge}_{\triangledown a}(\beta\alpha)}(\rho_0))+\\ &\ \ \ \ \ \ \ \ \ \ \ \ \ \ \ \ \ \ \ \ \ \ \ d_a(\mathcal{E}_{{\tt purge}_{\triangledown a}(\beta(b,c)\alpha)}(\rho_0),\mathcal{E}_{\beta(b,c)\alpha}(\rho_0))\\ &\leq 2\cdot K(\mathbb{S},\leadsto)=0.\end{split}\end{equation*} Consequently, it holds that $\rho\stackrel{a}{\sim}\mathcal{E}_{b,c}(\rho)$. \hfill$\blacksquare$

\subsubsection*{A.3. Proof of Theorem \ref{unw2}}

By definition, we only need to prove that for every agent $a\in A$ and for all action sequence $\alpha\in (A\times C)^\ast$ with $|\alpha|\leq t$, 
$$d_a(\mathcal{E}_\alpha (\rho_0),\mathcal{E}_{{\tt purge}_{\triangledown a}(\alpha)}(\rho_0))\leq \epsilon_o+t\cdot \max\{\epsilon_s,\epsilon_l\}.$$ 
It follows from the approximate observation consistency that  \begin{equation*}\begin{split}d_a(\mathcal{E}_\alpha (\rho_0),\ &\mathcal{E}_{{\tt purge}_{\triangledown a}(\alpha)}(\rho_0))\\ &\leq \delta_a(\mathcal{E}_\alpha (\rho_0),\mathcal{E}_{{\tt purge}_{\triangledown a}(\alpha)}(\rho_0))+\epsilon_o.\end{split}\end{equation*}
Thus, it suffices to show that $$\delta_a(\mathcal{E}_\alpha (\rho_0),\mathcal{E}_{{\tt purge}_{\triangledown a}(\alpha)}(\rho_0))\leq t\cdot \max\{\epsilon_s,\epsilon_l\}.$$
We proceed by induction on the length $|\alpha|$ of $\alpha$. The basis case of $|\alpha|=0$ is clear. Now we assume that $\alpha=\alpha^\prime (b,c)$ and consider the following two cases:

Case 1. $b\leadsto a$. Then $${\tt purge}_{\triangledown a}(\alpha)={\tt purge}_{\triangledown a}(\alpha^\prime) (b,c),$$ and by the induction hypothesis on $\alpha^\prime$ we have:  
$$\delta_a(\mathcal{E}_{\alpha^\prime}(\rho_0),\mathcal{E}_{{\tt purge}_{\triangledown a}(\alpha^\prime)}(\rho_0))\leq (t-1)\cdot\max\{\epsilon_s,\epsilon_l\}$$ because $|\alpha^\prime|=|\alpha|-1\leq t-1$. Thus, by the approximate step consistency we obtain: \begin{equation*}\begin{split}
\delta_a(\mathcal{E}_{\alpha}&(\rho_0),\ \mathcal{E}_{{\tt purge}_{\triangledown a}(\alpha)}(\rho_0))\\ &= \delta_a(\mathcal{E}_{b,c}(\mathcal{E}_{\alpha}(\rho_0)),
\mathcal{E}_{b,c}(\mathcal{E}_{{\tt purge}_{\triangledown a}(\alpha)}(\rho_0)))\\ 
&\leq \delta_a(\mathcal{E}_{\alpha^\prime}(\rho_0),\mathcal{E}_{{\tt purge}_{\triangledown a}(\alpha^\prime)}(\rho_0))+\epsilon_s\\ &\leq (t-1)\cdot\max\{\epsilon_s,\epsilon_l\}+\epsilon_s\\ &\leq t\cdot\max\{\epsilon_s,\epsilon_l\}.
\end{split}\end{equation*}

Case 2. $b\not\leadsto a$. Then $${\tt purge}_{\triangledown a}(\alpha)={\tt purge}_{\triangledown a}(\alpha^\prime),$$ and the approximate local respect of $\leadsto$ and the induction hypothesis on $\alpha^\prime$ yield  
\begin{equation*}\begin{split}
\delta_a(\mathcal{E}_{\alpha}(&\rho_0),\mathcal{E}_{{\tt purge}_{\triangledown a}(\alpha)}(\rho_0))\\ &=\delta_a(\mathcal{E}_{b,c}(\mathcal{E}_{\alpha^\prime}(\rho_0)),
\mathcal{E}_{{\tt purge}_{\triangledown a}(\alpha^\prime)}(\rho_0))\\ 
&\leq \delta_a(\mathcal{E}_{b,c}(\mathcal{E}_{\alpha^\prime}(\rho_0)),
\mathcal{E}_{\alpha^\prime}(\rho_0)))\\ 
& \ \ \ \ \ \ \ \ \ \ \ \ \ \ \ \ + \delta_a(\mathcal{E}_{\alpha}(\rho_0)),\mathcal{E}_{{\tt purge}_{\triangledown a}(\alpha^\prime)}(\rho_0))\\ 
&\leq \epsilon_l+(t-1)\cdot\max\{\epsilon_s,\epsilon_l\}\\ &\leq t\cdot\max\{\epsilon_s,\epsilon_l\}.\end{split}\end{equation*} \hfill$\blacksquare$

\subsection*{A.4. Proof of Theorem~\ref{unw3}}

For each $a\in A$, and for any reachable density operators $\rho,\sigma$, we define: $$\delta_a(\rho,\sigma)=\sup_{\alpha\in (A\times C)^\ast}d_a(\mathcal{E}_\alpha(\rho),\mathcal{E}_\alpha(\sigma)).$$ It is easy to see that $\delta_a$ is a pseudo-distance for each $a\in A$. Step and Observation consistency follow immediately from the definition of $\delta_a$. We now prove the bounded local respect of $\leadsto$. It suffices to show that for any $a,b\in A$, $c\in C$, and reachable density operator $\rho$, if $b\not\leadsto a$, then $$\delta_a(\rho,\mathcal{E}_{b,c}(\rho))\leq 2K(\mathbb{S},\leadsto).$$ 
In fact, since $\rho$ is reachable, it holds that $\rho=\mathcal{E}_\beta(\rho_0)$ for some $\beta\in (A\times C)^\ast$. Thus, we have:\begin{equation*}\begin{split}
&\delta_a(\rho,\mathcal{E}_{b,c}(\rho))=\sup_{\alpha\in (A\times C)^\ast}d_a(\mathcal{E}_\alpha(\rho),\mathcal{E}_\alpha(\mathcal{E}_{b,c}(\rho)))\\ &=\sup_{\alpha\in (A\times C)^\ast}d_a(\mathcal{E}_{\beta\alpha}(\rho_0),\mathcal{E}_{\beta(b,c)\alpha}(\rho_0))\\ &\leq \sup_{\alpha\in (A\times C)^\ast}[d_a(\mathcal{E}_{\beta\alpha}(\rho_0),\mathcal{E}_{{\tt purge}_{\triangledown a}(\beta\alpha)}(\rho_0))\\ &\ \ \ \ \ \ \ \ \ \ \ \ \ \ \ \ \ \ \ \ \ \ +d_a(\mathcal{E}_{{\tt purge}_{\triangledown a}(\beta\alpha)}(\rho_0),\mathcal{E}_{\beta(b,c)\alpha}(\rho_0))]\\ &\leq \sup_{\alpha\in (A\times C)^\ast}d_a(\mathcal{E}_{\beta\alpha}(\rho_0),\mathcal{E}_{{\tt purge}_{\triangledown a}(\beta\alpha)}(\rho_0))\\ &\ \ \ \ \ \ \ \ \ \ \ \ +\sup_{\alpha\in (A\times C)^\ast}d_a(\mathcal{E}_{{\tt purge}_{\triangledown a}(\beta\alpha)}(\rho_0),\mathcal{E}_{\beta(b,c)\alpha}(\rho_0))\\ &= \sup_{\alpha\in (A\times C)^\ast}d_a(\mathcal{E}_{\beta\alpha}(\rho_0),\mathcal{E}_{{\tt purge}_{\triangledown a}(\beta\alpha)}(\rho_0))\\ &\ \ \ \ \ \ \ \ \ +\sup_{\alpha\in (A\times C)^\ast}d_a(\mathcal{E}_{{\tt purge}_{\triangledown a}(\beta (b,c)\alpha)}(\rho_0),\mathcal{E}_{\beta(b,c)\alpha}(\rho_0))\\ &\leq 2K(\mathbb{S},\leadsto)
\end{split}\end{equation*} because $b\not\leadsto a$ implies ${\tt purge}_{\triangledown a}(\beta (b,c)\alpha)={\tt purge}_{\triangledown a}(\beta\alpha). \hfill\blacksquare$

\subsection*{B. Proof of Theorems in Section~\ref{COMPS}}

We first present two technical lemmas needed in the proofs of Theorems~\ref{comp-sec0} and \ref{Comp-SEC}. 

\textit{Lemma B.1:}
If all POVM measurements in $\mathbb{M}$ are all performed on the first subsystem, then we have: $$d_\mathbb{M}(\rho_1\otimes\rho,\rho_2\otimes\rho^\prime)=d_\mathbb{M}(\rho_1,\rho_2)$$
 for any density operators $\rho_1,\rho_2$ of the first subsystem, and for any density operators $\rho, \rho^\prime$ of the second subsystem.
 
 \textit{Proof:} For each POVM measurement $E$ in $\mathbb{M}$, since it is performed only on the first subsystem, we can write $E=\{E_\lambda\otimes I\}_\lambda$, where $E_\lambda$ is an operator on the first subsystem for every $\lambda$, and $I$ is the identity operator on the second subsystem. Thus, \begin{equation*}\begin{split}
p_E(\rho_1\otimes\rho,\lambda)&=tr((E_\lambda\otimes I)(\rho_1\otimes\rho))\\ &=tr((E_\lambda\rho_1)\otimes\rho)=tr(E_\lambda\rho_1)\cdot tr(\rho)\\ &=tr(E_\lambda\rho_1)=p_E(\rho_1,\lambda);
 \end{split}\end{equation*} that is, the probability distribution defineed by $E$ and $\rho_1\otimes\rho$ is equal to that defined by $E$ and $\rho_1$. Similarly, we have $p_E(\rho_2\otimes\rho^\prime)=p_E(\rho_2)$. Therefore, $$d(p_E(\rho_1\otimes\rho),p_E(\rho_2\otimes\rho^\prime))=d(p_E(\rho_1),p_E(\rho_2)),$$ and the conclusion follows.\hfill $\blacksquare$

\textit{Lemma B.2:} (Convexity of Measurement Distance) Let $\mathbb{M}$ be a family of POVM measurements, let $\{p_i\}$ be a probability distribution, and let $\rho_i$ and $\sigma_i$ be density operators for every $i$. Then $$d_\mathbb{M}(\sum_ip_i\rho_i,\sum_ip_i\sigma_i)\leq\sum_i p_i d_{\mathbb{M}}(\rho_i,\sigma_i).$$

\textit{Proof:} We first prove the conclusion in the special case where $\mathbb{M}$ is a singleton $\{E\}$. In thos case, we simply write $d_E$ for $d_\mathbb{M}$. Suppose that $E=\{E_\lambda\}_{\lambda\in\Lambda}$. By definition, we have:\begin{equation}\label{specas}\begin{split}&d_E(\sum_ip_i\rho_i,\sum_ip_i\sigma_i)\\ &=\sum_{\lambda\in\Lambda}|tr(E_\lambda(\sum_ip_i\rho_i))-tr(E_\lambda(\sum_ip_i\sigma_i))|\\ &=\sum_{\lambda\in\Lambda}|\sum_i p_i tr(E_\lambda\rho_i)-\sum_i p_i tr(E_\lambda\sigma_i)|
\\ &\leq \sum_{\lambda\in\Lambda}\sum_i p_i| tr(E_\lambda\rho_i)-tr(E_\lambda\sigma_i)|\\ &= \sum_i p_i \sum_{\lambda\in\Lambda}|tr(E_\lambda\rho_i)-tr(E_\lambda\sigma_i)|
\\ &= \sum_i p_i d_E(\rho_i,\sigma_i).
\end{split}\end{equation} 

In general, it follows from Eq.~(\ref{specas}) that \begin{equation*}\begin{split}
d_\mathbb{M}(\sum_ip_i\rho_i,\sum_ip_i\sigma_i)&=\sup_{E\in\mathbb{M}}d_E(\sum_ip_i\rho_i,\sum_ip_i\sigma_i)\\ &\leq \sup_{E\in\mathbb{M}}\sum_i p_i d_E(\rho_i,\sigma_i)\\ 
&\leq \sum_i p_i \sup_{E\in\mathbb{M}}d_E(\rho_i,\sigma_i)\\ &=\sum_i p_i d_\mathbb{M}(\rho_i,\sigma_i).
\end{split}\end{equation*} \hfill$\blacksquare$

Now we are ready to prove Theorems~\ref{comp-sec0} and \ref{Comp-SEC}. Theorem~\ref{comp-sec0} can be proved in a way that is similar to but easier than the proof of Theorem~\ref{Comp-SEC}. 

For Theorem~\ref{Comp-SEC}, we only prove 1) because 2) can be proved in the same way. For each $a\in A\cup A^\prime$, we put $$\triangledown a=\{b\in A\cup A^\prime:{\rm not}\ b(\leadsto\cup\leadsto^\prime) a\}.$$ Then it suffices to show that for all $\alpha\in [(A\cup A^\prime)\times (C\cup C^\prime)]^\ast$, 
\begin{equation}\begin{split}D_a(\mathcal{F}_\alpha(\sigma_0), \ &\mathcal{F}_{{\tt purge}_{\triangledown a}(\alpha)}(\sigma_0)\\ &\ \ \ \ \ \ \ \ \leq SK(\mathbb{S},\leadsto)+SK(\mathbb{S}^\prime,\leadsto^\prime),\end{split}\end{equation} where $D_a$ is the measurement distance in the composed system $\mathbb{T}$; that is, $D_a=d_{\mathbb{N}_a}$. 
Let $\beta,\gamma,\delta$ be the subsequences of $\alpha$ consisting of elements in $A\times C, A^\prime\times C^\prime$ and $[(A\setminus A^\prime)\times C^\prime]\cup[(A^\prime\setminus A)\times C]$, respectively, and let $\beta^\prime,\gamma^\prime,\delta^\prime$ be the corresponding subsequences of ${\tt purge}_{\triangledown a}(\alpha).$  Since the initial state $\sigma_0$ is separable, we can write $\sigma_0$ in the following way: $$\sigma_0=\sum_i p_i(\rho_i\otimes\rho_i^\prime),$$ where $\{p_i\}$ is a probability distribution, and $\rho_i, \rho_i^\prime$ are density operators in $\mathcal{H}$ and $\mathcal{H}^\prime$, respectively, for every $i$. 

(i) By definition, we obtain: $$\rho_0=tr_{\mathcal{H}^\prime}(\sigma_0)=\sum_ip_i\rho_i.$$ So, $\rho_0$ is a mixture of ensemble $\{(p_i,\rho_i)\}$. Similarly, we see that $\rho_0^\prime$ is a mixture of ensemble $\{(p_i,\rho_i^\prime)\}$.

(ii) It follows from Eq.~(\ref{comp1}) that 
\begin{equation*}\begin{split}\mathcal{F}_\alpha(\sigma_0)&=\sum_ip_i\mathcal{F}_\alpha(\rho_i\otimes\rho_i^\prime)\\ &=\sum_ip_i[\mathcal{E}_\beta(\rho_i)\otimes\mathcal{E}_\gamma^\prime(\rho_i^\prime)]\end{split}\end{equation*} because any operator of the form $\mathcal{E}\otimes\mathcal{I}_{\mathcal{H}^\prime}$ commutes with any operator of the form $\mathcal{I}_{\mathcal{H}}\otimes\mathcal{E}^\prime$. Similarly, we have:  
\begin{equation*}\mathcal{F}_{{\tt purge}_{\triangledown a}(\alpha)}(\sigma_0)=\sum_ip_i[\mathcal{E}_{\beta^\prime}(\rho_i)\otimes\mathcal{E}_{\gamma^\prime}^\prime(\rho_i^\prime)].
\end{equation*}

Now we consider the following three cases:

Case 1. $a\in A\setminus A^\prime$. We write: $$\triangledown_1 a=\{b\in A:b\not\leadsto a\}.$$ Then by the compatibility of $\leadsto$ and $\leadsto^\prime$ we have: 
\begin{equation*}\begin{split}\triangledown a&=\triangledown_1 a\cup (A^\prime\setminus A),\\ 
\beta^\prime&={\tt purge}_{\triangledown_1 a}(\beta),\\ 
\gamma^\prime&=\epsilon\ ({\rm empty\ string}).\end{split}\end{equation*} Furthermore, using Eq.~(\ref{comp2}) and Lemmas B.1 and B.2 we obtain:
\begin{equation*}\begin{split}&D_a(\mathcal{F}_\alpha(\sigma_0), \mathcal{F}_{{\tt purge}_{\triangledown a}(\alpha)}(\sigma_0))\\
&=d_{\mathbb{M}_a}(\sum_ip_i[\mathcal{E}_\beta(\rho_i)\otimes\mathcal{E}_\gamma^\prime(\rho_i^\prime)],\sum_ip_i[\mathcal{E}_{\beta^\prime}(\rho_i)\otimes\mathcal{E}^\prime_{\gamma^\prime}(\rho_i^\prime)])\\
&=d_{\mathbb{M}_a}(\sum_ip_i[\mathcal{E}_\beta(\rho_i)\otimes\mathcal{E}_\gamma^\prime(\rho_i^\prime)],\\ &\ \ \ \ \ \ \ \ \ \ \ \ \ \ \ \ \ \ \ \ \ \ \ \ \ \ \ \ \sum_ip_i[\mathcal{E}_{{\tt purge}_{\triangledown_1 a}(\beta)}(\rho_i)\otimes\rho_i^\prime)])\\ 
&\leq\sum_ip_i d_{\mathbb{M}_a}(\mathcal{E}_\beta(\rho_i)\otimes\mathcal{E}_\gamma^\prime(\rho_i^\prime),\mathcal{E}_{{\tt purge}_{\triangledown_1 a}(\beta)}(\rho_i)\otimes\rho_i^\prime))\\ 
&\leq\sum_ip_i d_{\mathbb{M}_a}(\mathcal{E}_\beta(\rho_i),\mathcal{E}_{{\tt purge}_{\triangledown_1 a}(\beta)}(\rho_i))\\ &
\leq\sum_ip_iK(\mathbb{S}[\rho_i],\leadsto)\\ 
&\leq SK(\mathbb{S},\leadsto).
\end{split}\end{equation*}

Case 2. $a\in A^\prime\setminus A$. Similar to Case 1.

Case 3. $a\in A\cap A^\prime$. We write: $$\triangledown_2 a=\{b\in A^\prime:b\not\leadsto^\prime a\}.$$ Then by the compatibility of $\leadsto$ and $\leadsto^\prime$ we have: 
\begin{equation*}\begin{split}
\triangledown a&=\triangledown_1 a\cup\triangledown_2 a,\\ \beta^\prime&={\tt purge}_{\triangledown_1 a}(\beta),\\ 
\gamma^\prime&={\tt purge}_{\triangledown_2 a}(\gamma).\end{split}\end{equation*}
It follows from Eq.~(\ref{comp2}) and Lemma B.1 that 
\begin{equation*}\begin{split}&D_a(\mathcal{E}_\beta(\rho_i)\otimes\mathcal{E}^\prime_\gamma(\rho_i^\prime),\mathcal{E}_{{\tt purge}_{\triangledown_1 a}(\beta)}(\rho_i)\otimes\mathcal{E}^\prime_\gamma(\rho_i^\prime))\\ &=d_{\mathbb{M}_a\cup\mathbb{M}^\prime_a}(\mathcal{E}_\beta(\rho_i)\otimes\mathcal{E}^\prime_\gamma(\rho_i^\prime),\mathcal{E}_{{\tt purge}_{\triangledown_1 a}(\beta)}(\rho_i)\otimes\mathcal{E}^\prime_\gamma(\rho_i^\prime))\\
&=\max\{d_{\mathbb{M}_a}(\mathcal{E}_\beta(\rho_i)\otimes\mathcal{E}^\prime_\gamma(\rho_i^\prime),\mathcal{E}_{{\tt purge}_{\triangledown_1 a}(\beta)}(\rho_i)\otimes\mathcal{E}^\prime_\gamma(\rho_i^\prime)),\\ &\ \ \ \ \ \ \ \ \ \ \ d_{\mathbb{M}^\prime_a}(\mathcal{E}_\beta(\rho_i)\otimes\mathcal{E}^\prime_\gamma(\rho_i^\prime),\mathcal{E}_{{\tt purge}_{\triangledown_1 a}(\beta)}(\rho_i)\otimes\mathcal{E}^\prime_\gamma(\rho_i^\prime))\}\\
&=d_{\mathbb{M}_a}(\mathcal{E}_\beta(\rho_i),\mathcal{E}_{{\tt purge}_{\triangledown_1 a}(\beta)}(\rho_i))\\
&\leq K(\mathbb{S}[\rho_i],\leadsto).
\end{split}\end{equation*}Similarly, we have: 
\begin{equation*}\begin{split}D_a(\mathcal{E}_{{\tt purge}_{\triangledown_1 a}(\beta)}(\rho_i)&\otimes\mathcal{E}^\prime_\gamma(\rho_i^\prime),\mathcal{E}_{{\tt purge}_{\triangledown_1 a}(\beta)}(\rho_i)\\ &\otimes\mathcal{E}^\prime_{{\tt purge}_{\triangledown_2 a}(\gamma)}(\rho_i^\prime))
\leq K(\mathbb{S}^\prime[\rho_i^\prime],\leadsto^\prime).
\end{split}\end{equation*}Therefore, it holds that 
\begin{equation*}\begin{split}
&D_a(\mathcal{E}_\beta(\rho_i)\otimes\mathcal{E}^\prime_\gamma(\rho_i^\prime), \mathcal{E}_{{\tt purge}_{\triangledown_1 a}(\beta)}(\rho_i)\otimes\mathcal{E}^\prime_{{\tt purge}_{\triangledown_2 a}(\gamma)}(\rho_i^\prime))\\
&\leq D_a(\mathcal{E}_\beta(\rho_i)\otimes\mathcal{E}^\prime_\gamma(\rho_i^\prime),\mathcal{E}_{{\tt purge}_{\triangledown_1 a}(\beta)}(\rho_i)\otimes\mathcal{E}^\prime_\gamma(\rho_i^\prime))
\\ &\ \ \ \ \ \ \ \ \ \ +D_a(\mathcal{E}_{{\tt purge}_{\triangledown_1 a}(\beta)}(\rho_i)\otimes\mathcal{E}^\prime_\gamma(\rho_i^\prime),\\ &\ \ \ \ \ \ \ \ \ \ \ \ \ \ \ \ \ \ \ \ \mathcal{E}_{{\tt purge}_{\triangledown_1 a}(\beta)}(\rho_i)\otimes\mathcal{E}^\prime_{{\tt purge}_{\triangledown_2 a}(\gamma)}(\rho_i^\prime))\\ 
&\leq K(\mathbb{S}[\rho_i],\leadsto)+K(\mathbb{S}^\prime[\rho_i^\prime],\leadsto^\prime).
\end{split}\end{equation*} Finally, by Lemma B.2 we obtain:
\begin{equation*}\begin{split}&D_a(\mathcal{F}_\alpha(\sigma_0), \mathcal{F}_{{\tt purge}_{\triangledown a}(\alpha)}(\sigma_0))\\
&=D_a(\sum_ip_i[\mathcal{E}_\beta(\rho_i)\otimes\mathcal{E}_\gamma^\prime(\rho_i^\prime)],\\ & \ \ \ \ \ \ \ \ \ \ \ \ \sum_ip_i[\mathcal{E}_{{\tt purge}_{\triangledown_1 a}(\beta)}(\rho_i)\otimes\mathcal{E}^\prime_{{\tt purge}_{\triangledown_1 a}(\gamma)}(\rho_i^\prime))])\\ &\leq \sum_ip_iD_a(\mathcal{E}_\beta(\rho_i)\otimes\mathcal{E}_\gamma^\prime(\rho_i^\prime),\\ & \ \ \ \ \ \ \ \ \ \ \ \ \ \ \ \ \ \ \ \ \ \ \mathcal{E}_{{\tt purge}_{\triangledown_1 a}(\beta)}(\rho_i)\otimes\mathcal{E}^\prime_{{\tt purge}_{\triangledown_1 a}(\gamma)}(\rho_i^\prime)))\\ &\leq\sum_ip_i[K(\mathbb{S}[\rho_i],\leadsto)+K(\mathbb{S}^\prime[\rho_i^\prime],\leadsto^\prime)]\\ &= \sum_ip_iK(\mathbb{S}[\rho_i],\leadsto)+\sum_ip_iK(\mathbb{S}^\prime[\rho_i^\prime],\leadsto^\prime)\\ &\leq SK(\mathbb{S},\leadsto)+SK(\mathbb{S}^\prime,\leadsto^\prime)]
\end{split}\end{equation*} \hfill$\blacksquare$

\subsection*{C. Proof of Theorem~\ref{main-acc}}

We first prove the following two claims:

$${\rm \underline{Claim 1}}.\ \ \ \ \delta_a(\mathcal{E}_{b,c}(\rho),\mathcal{E}_{b,c}(\sigma))\leq\delta_a(\rho,\sigma)+2\epsilon.\ \ \ \ \ \ \ \ $$

To prove this claim, we only need to show that for any $K\in read(a)$, $$d(tr_K(\mathcal{E}_{b,c}(\rho)),tr_K(\mathcal{E}_{b,c}(\sigma)))\leq\delta_a(\rho,\sigma)+2\epsilon.$$ We consider the following cases:

Case 1. $\rho$ and $\mathcal{E}_{b,c}(\rho)$ are $\epsilon-$discriminable on $K$. Then it follows from (RM2) that $\mathcal{E}_{b,c}(\rho)$ and $\mathcal{E}_{b,c}(\sigma)$ are not $\delta_b(\rho,\sigma)-$discriminable; that is, \begin{equation}\label{acm-p}d(tr_K(\mathcal{E}_{b,c}(\rho)),tr_K(\mathcal{E}_{b,c}(\sigma)))\leq\delta_b(\rho,\sigma).\end{equation} On the other hand, by (RM3) we have $K\cap L\neq\emptyset$ for some $L\in alter(b)$, and by condition 2) of Definition~\ref{acm-s} and $K\in read(a)$ we further obtain $b\leadsto a$. This together with condition 1) of Definition~\ref{acm-s} implies that $read(b)\subseteq read(a)$, and by definition we have $\delta_b(\rho,\sigma)\leq\delta_a(\rho,\sigma)$. Therefore, it follows from Eq.~(\ref{acm-p}) that $$d(tr_K(\mathcal{E}_{b,c}(\rho)),tr_K(\mathcal{E}_{b,c}(\sigma)))\leq\delta_a(\rho,\sigma).$$

Case 2. $\sigma$ and $\mathcal{E}_{b,c}(\sigma)$ are $\epsilon-$discriminable on $K$. Similar to Case 1.

Case 3. $\rho$ and $\mathcal{E}_{b,c}(\rho)$ are not $\epsilon-$discriminable on $K$, and $\sigma$ and $\mathcal{E}_{b,c}(\sigma)$ are not $\epsilon-$discriminable on $K$. Then it holds that \begin{equation*}\begin{split}&d(tr_K(\mathcal{E}_{b,c}(\rho)),tr_K(\mathcal{E}_{b,c}(\sigma)))\leq d(tr_K(\mathcal{E}_{b,c}(\rho)),tr_K(\rho))\\ &\ \ \ \ \ \ \ \ \ \ +d(tr_K(\rho),tr_K(\sigma))+d(tr_K(\sigma),tr_K(\mathcal{E}_{b,c}(\sigma)))\\ &\leq\delta_a(\rho,\sigma)+2\epsilon.
\end{split}\end{equation*}

$${\rm \underline{Claim 2}}. \ \ \ \ b\not\leadsto a\Rightarrow\delta_a(\rho,\mathcal{E}_{b,c}(\rho))\leq\epsilon.\ \ \ \ \ \ \ \ \ \ \ \ \ \ \ \ \ $$

To prove this claim, we only need to show that for any $K\in read(a)$, $$d(tr_K(\rho),tr_K(\mathcal{E}_{b,c}(\rho))\leq\epsilon.$$ This can be done by refutation. If there exists $K\in read(a)$ such that 
$$d(tr_K(\rho),tr_K(\mathcal{E}_{b,c}(\rho))>\epsilon;$$ that is, $\rho$ and $\mathcal{E}_{b,c}(\rho)$ are $\epsilon-$discriminable on $K$, then by (RM3) we assert that there exists $L\in alter(a)$ with $K\cap L\neq\emptyset$. It follows from condition 2) of Definition~\ref{acm-s} that $b\leadsto a$. This contradicts to the assumption that $b\not\leadsto a$.

Finally, by combining (RM1) and Claims 1 and 2 and applying Theorem~\ref{unw2} we obtain: $$K_t(\mathbb{S},\leadsto)\leq\theta+t\cdot\max\{2\epsilon,\epsilon\}=\theta+2t\epsilon.$$ \hfill$\blacksquare$
\end{document}